\documentclass[prd,preprint,showpacs]{revtex4}%
\usepackage{amssymb}
\usepackage{amsmath}
\usepackage{graphicx}
\usepackage{bm}
\usepackage{dcolumn}%
\usepackage{amsfonts}
\providecommand{\U}[1]{\protect\rule{.1in}{.1in}}
\begin{document}
\title{From scale properties of physical amplitudes to a predictive formulation of
the Nambu--Jona-Lasinio model.}
\author{O.A. Battistel$^{1}$and G. Dallabona$^{2}$ }
\affiliation{$^{1}$Departamento de F\'{\i}sica, Universidade Federal de Santa Maria,
97119-900 Santa Maria, RS, Brazil}
\affiliation{$^{2}$Departamento de Ci\^{e}ncias Exatas, Universidade Federal de Lavras,}
\affiliation{Cx. Postal 37, 37200-000, Lavras, MG, Brazil}

\begin{abstract}
The predictive power of the NJL model is considered in the light of a novel
strategy to handle the divergences typical of perturbative calculations. The
referred calculational strategy eliminates unphysical dependencies on the
arbitrary choices for the routing of internal momenta and symmetry violating
terms. In the present work we extend a previous one on the same issue by
including vector interactions and performing the discussion in a more general
context: it is considered the role of scale arbitrariness for the consistency
of the calculations. We show that the imposition of arbitrary scale
independence for the consistent regularized amplitudes lead to additional
properties for the irreducible divergent objects. These properties allow us to
parametrize the remaining freedom in terms of a unique constant where resides
all the arbitrariness involved. By searching for the best value for the
arbitrary parameter we find a critical condition for the existence of an
acceptable physical value for the dynamically generated quark mass. Such
critical condition fixes the remaining arbitrariness turning the NJL into a
predictive model in the sense that its phenomenological consequences do not
depend on possible choices made in intermediary steps. Numerical results are
obtained for physical quantities like the vector and axial-vector masses and
their coupling constants as genuine predictions.

\end{abstract}
\maketitle

\section{Introduction}

It would be desirable to describe all the phenomenology of interacting quarks
from the point of view of quantum chromodynamics (QCD), the fundamental
(renormalizable) theory of strong interactions. Even if this is in principle
possible, there are many difficulties to be overcome such that effective
models, having the main symmetries of the theory, have been frequently used to
describe some aspects of the QCD phenomenology. In this context the
Nambu--Jona-Lasinio (NJL) model \cite{Nambu} is of indisputable popularity.
The model has been used to describe the low energy hadronic phenomenology
through observables like hadronic masses spectrum, correlation and structure
functions in vacuum since the downing of QCD. Such observables, among others,
have been considered at finite densities and temperatures, which stated a
large lack of phenomenological aspects associated with systems of quarks and
gluons-- for a representative list of references see the reviews in Refs.
\cite{VW, Klevansky, HK-rev, Bochum, Bijnens, Bernard, Ripka, Bub}. The model,
from the point of view of quantum field theory (QFT) formalism, is a
nonrenormalizable theory, due to the presence of four (1/2 spin) fermion
interactions in the model Lagrangian, such that predictions only make sense
within the context of a particular level of approximation. Having this
limitation in mind, an uncountable number of valuable works have been and are
still being produced nowadays. There are many reasons for such interest. Two
of them, however, deserve special attention. The first one is the general
acceptance that the NJL model captures many of the essential features of
chiral symmetry in QCD. In the limit of exact chiral symmetry the fermions of
the model are massless and the interaction Lagrangian density contains the
chirally symmetric combinations of four-fermion interactions. The second
attractive feature is the fact that the model realizes the dynamically
breaking of chiral symmetry already at the one-loop (mean field) approximation
such that the fermions become massive. The difficulties with the model
predictions are related to the nonrenormalizable character of the
corresponding relativistic theory such that they are unavoidable intimately
compromising with the specific strategy adopted to handle the ultraviolet
divergences present in the amplitudes in each particular level of
approximation. Consequently, in order to extract physical predictions, we have
to specify a procedure for the necessary handling of the divergent amplitudes
as the first step. Concerning this aspect this is not different from the
procedure adopted in any relativistic QFT. However, while in a renormalizable
theory all the parameters alien to the theory which are introduced in the
regularization of the amplitudes, can be completely removed from the physical
amplitudes after the divergences are isolated and eliminated through the
reparametrization of the theory, this cannot be done in a nonrenormalizable
model. This means that the parametrization of the model, which enables us to
remove the divergences, must be made in a particular way restricted to the
particular level of approximation considered. Within this context, in general,
practitioners of the NJL model have adopted the attitude of using it as being
a regularization-dependent model, including the regularization procedure as a
part of the definition of the model.

The regularization of divergent amplitudes, on the other hand, is a dangerous
process since it involves many types of arbitrariness in the manipulation of
improper integrals. They may be converted into ambiguities if the results
become dependent on the choices involved. Among such ambiguities there are
those associated with the arbitrary routing of the momenta in internal lines
of divergent loop amplitudes, whose existence is invariably associated with
the violation of space-time homogeneity. Another kind of ambiguities with
important consequences in the present contribution is the one associated with
the choice of the common scale for the divergent and finite parts of
amplitudes that lead to the breaking of scale invariance. In general,
ambiguous terms lead to violations of symmetry relations of global and local
gauge symmetries. The most commonly used regularization procedures, within the
context of NJL model, such as three-and four-momentum cutoff, Pauli-Villars
and proper-time lead to one or more of such symmetry violations
\cite{Bron,Doring,DavERA}. Dimensional regularization (DR) \cite{DR}, on the
other hand, leads to ambiguities-free and symmetry-preserving amplitudes but
it is not adequate to NJL model due to the fact that the quadratic divergence
which appears in almost all one-loop amplitudes must be assumed as zero in the
zero-mass limit, which is associated with the chiral symmetry restoration (at
high densities and temperatures). Such type of difficulties lead Willey
\cite{Willey} and Gherghetta \cite{Gherghetta} to conclude that there is no
way to make consistent physical predictions with the NJL model using
traditional regularization techniques. However, this question was considered
in a later work by Battistel and Nemes \cite{Orimar-Nemes} by using a novel
strategy to handle the divergent amplitudes \cite{Orimar-tese}. The referred
investigation, within the context of the gauged NJL model revealed that it is
possible to obtain ambiguities-free and symmetry-preserving amplitudes, which
is the first and crucial obstacle to be removed in order to get a predictive
model. We said the first obstacle due to the fact that the requirements found
as necessary properties for consistent regularizations, in order to remove, in
a systematic way, the potentially ambiguous and symmetry violating terms, do
not fix a single complete regularization. The restrictions imposed, in fact,
state only a class of regularizations. This is not sufficient to make a
nonrenormalizable model predictive because in this case the model predictions,
within the context of a particular approximation scheme, may be sensitive to
the particular aspects of the regularization employed, even if such
regularization belongs to the class of consistent ones. In this scenario the
amplitudes are symmetry-preserving and free from ambiguities associated with
the choices of routings for the internal lines momenta but the predictions
still involves an arbitrariness which is related to the choice of the
particular regularization. The practical consequence is that additional
phenomenological information must be used in order to parametrize the model.
Such input invariably belongs to the phenomenological scope of the model i.e.
it is a part of the phenomenology which we want to describe through the model
considered. In the context of NJL model, it is common to use the pion mass to
select the value of the cutoff $\Lambda$. The pion mass, on the other hand, is
a low energy hadronic observable which we would like to predict from the quark
properties present at the model Lagrangian. This situation is very frustrating
and deserves additional investigations in order to be solved, making the model
predictive, which means to obtain its consequences in a way completely
independent of the choices involved in intermediary steps of the calculations.

Having this in mind, in a recent contribution, the question of predictability
of the NJL model has been again considered \cite{Orimar-PRD-NJL}. In a
complete and detailed investigation, where even the amplitudes having tensor
operators have been considered, it was stated that the amplitudes can be
obtained preserving their symmetry relations as well as freeing them from the
dependences on the choices for the internal lines momenta routing through the
identification of general and universal properties for the divergent Feynman
integrals, which we denominated consistency relations. An additional step has
been performed by using specific properties for the so called irreducible
divergent objects due to the fact that, following this procedure, the
remaining freedom associated with the choice of the regularization can be put
in terms of an arbitrary parameter. Given this situation we, in a first
moment, have to recognize that the parametrization of the model and,
consequently, the predictions, are really regularization dependent. However,
when we search for the best value for the arbitrary parameter, looking for the
value of the dynamically generated quark mass, we identified the existence of
a critical condition. Only one value for the arbitrary parameter is associated
with a unique and real value for the quark mass. Given this fact we concluded
that the predictions of the NJL model are not dependent on choices and we
denominated this interpretation as a predictive formulation of the NJL model.
Within this formulation, the parametrization of the model requires only three
inputs which are chosen as the value for the quark condensate, the pion decay
constant and the current quark mass. All the predictions, including the pion
mass, are made without additional parameters like the usual regularization
parameters (cutoffs). The numerical results obtained within the context of our
formulation are in good agreement with the phenomenological ones. This aspect
has been confirmed in an independent investigation made by Rochev
\cite{Rochev}.

Motivated by these surprising results, in the present work we would like to
take an additional step in the construction of the predictive formulation of
the NJL model, by introducing the most general kind of arbitrariness involved
in the perturbative evaluation of physical amplitudes which is the scale
arbitrariness. We will show that the scale properties of the physical
amplitudes are the adequate ingredient to discuss and understand the
consistency in perturbative calculations involving divergences. The scale
properties of the irreducible divergent objects are a natural consequence of
the imposition of scale independence in the manipulation and calculations
involving the intrinsic mathematical indefinitions associated with the
improper integrals typical of the perturbative calculations. In addition to
this aspect, in the present work, we consider a more general analysis of the
model parametrization by considering analytical solutions for the equation
which generate the critical condition for the dynamically quark mass which, in
addition to the more transparent analysis, allows us to obtain clean and sound
clarifications through the comparison of our results with those performed with
traditional treatments employing regularizations. The phenomenology of vector
mesons is also considered.

The work is organized in the following way: In the section II the model is
presented and the required definitions are introduced. In the section III we
consider the strategy we adopt to handle with the divergences and to explicit
the amplitudes which are necessary in the study of the considered
phenomenology of mesons presented in the section IV. Finally, in Section V we
make our final comments and conclusions.

\section{The model and definitions}

A very general Lagrangean constructed with a self interacting Dirac field
consistent with chiral symmetry (broken by the mass term) can be adopted as
having the form \cite{Jaminon-Arriola}:%
\begin{align*}
\mathcal{L} &  =\overline{\psi}\left(  x\right)  \left(  i\gamma^{\mu}%
\partial_{\mu}-m_{0}\right)  \psi\left(  x\right) \\
&  +\frac{G_{S}}{2}\left[  \left(  \overline{\psi}\left(  x\right)
\psi\left(  x\right)  \right)  ^{2}+\left(  \overline{\psi}\left(  x\right)
i\gamma_{5}\overrightarrow{\tau}\psi\left(  x\right)  \right)  ^{2}\right] \\
&  -\frac{G_{V}}{2}\left[  \left(  \overline{\psi}\left(  x\right)
\gamma_{\mu}\overrightarrow{\tau}\psi\left(  x\right)  \right)  ^{2}+\left(
\overline{\psi}\left(  x\right)  \gamma_{\mu}\gamma_{5}\overrightarrow{\tau
}\psi\left(  x\right)  \right)  ^{2}\right] \\
&  -\frac{G_{T}}{2}\left[  \left(  \overline{\psi}\left(  x\right)
\sigma_{\mu\nu}\overrightarrow{\tau}\psi\left(  x\right)  \right)
^{2}-\left(  \overline{\psi}\left(  x\right)  \sigma_{\mu\nu}\psi\left(
x\right)  \right)  ^{2}\right]  .
\end{align*}
If we take $\psi$ as the quark field ($u$ and $d$ flavors), $m_{0}$ as the
current quark mass ($m_{u}=m_{d}$), which explicitly breaks the chiral
symmetry, and $G_{S}$, $G_{V}$ and $G_{T}$ as the scalar (pseudo), vector
(axial) and tensor coupling strengths, respectively, then the above functional
represents the Lagrangian of the extended SU(2) NJL model. Through this model
it is possible to describe low energy hadronic observables. It is precisely
this challenge we will consider in what follows. In order to emphasize the
main aspects involved and to make the discussions transparent, in the present
contribution we adopt $G_{T}=0$. $\ $We start by introducing the required definitions.

The nonperturbative quark propagator $S(p)$ is given in terms of the
self-energy $\Sigma(p)$ as
\begin{equation}
S^{-1}(p)=\not p -\Sigma(p).\label{quark_prop}%
\end{equation}
In the mean field approximation, the self-energy is momentum independent
$\Sigma(p)\equiv M$, with $M$ satisfying a gap equation
\begin{equation}
M=m_{0}-2\,G_{S}N_{f}\left\langle \overline{\psi}\psi\right\rangle
,\label{gap_eq}%
\end{equation}
where $N_{f}=2$ is the number of flavors, and $\left\langle \overline{\psi
}\psi\right\rangle $ is the one-flavor, Lorentz scalar one-point function (the
quark condensate) given by
\begin{equation}
\left\langle \overline{\psi}\psi\right\rangle =N_{c}T^{S},\label{cond}%
\end{equation}
where $N_{c}=3$ is the number of colors and $T^{S}$ is the scalar one-point
amplitude defined by
\begin{equation}
T^{S}=-i\int\frac{d^{4}k}{\left(  2\pi\right)  ^{4}}t^{S},\ \ \ \ \ \ t^{S}%
=tr\left\{  \frac{1}{\left(  \not k +\not k _{1}\right)  -M}\right\}
.\label{T_S_def}%
\end{equation}
Here $k_{1}$\ is an arbitrary internal line momentum (for details please see
the next section).

\ In the context of NJL model, mesons are relativistic quark-antiquark bound
states which, in the random-phase approximation (RPA), its propagators can be
written as (see for example Ref.~\cite{Klevansky,HK-rev,Alkofer})
\begin{equation}
D_{\mathcal{M}_{1}\mathcal{M}_{2}}\left(  p^{2}\right)  =\frac{2G_{S}%
}{1-2G_{S}\Pi_{\mathcal{M}_{1}\mathcal{M}_{2}}\left(  p^{2}\right)
}\,,\label{mesons_prop}%
\end{equation}
where $\Pi_{\mathcal{M}_{1}\mathcal{M}_{2}}$ is the polarization functions
(fermion's loops) defined by
\begin{equation}
\Pi_{\mathcal{M}_{1}\mathcal{M}_{2}}\left(  p^{2}\right)  =-i\int\frac{d^{4}%
k}{\left(  2\pi\right)  ^{4}}\pi_{\mathcal{M}_{1}\mathcal{M}_{2}%
},\label{polarization}%
\end{equation}
where
\begin{equation}
\pi_{\mathcal{M}_{1}\mathcal{M}_{2}}=\mathrm{Tr}\left\{  \Gamma_{\mathcal{M}%
_{1}}S\left(  k+k_{1}\right)  \Gamma_{\mathcal{M}_{2}}S\left(  k+k_{2}\right)
\right\}  ,\label{integrand}%
\end{equation}
with $S$ being the quark propagator defined previously. $\Gamma_{\mathcal{M}}$
stands for the flavor and Dirac matrices giving the quantum numbers of the
meson $\mathcal{M}$. For example, for the neutral pion, $\Gamma_{\mathcal{M}%
}=\tau_{3}\gamma_{5}$, for the scalar-isoscalar meson, $\Gamma_{\mathcal{M}%
}=1$. In writing the equations above, we assumed the most general labels for
the momenta $k_{1}$ and $k_{2}$ running in the internal lines of the loop
integral. The physical momentum $p$ is defined as the difference $k_{1}-k_{2}$
as imposed by energy-momentum conservation at each vertex.

The approach used to study the mesonic phenomenology in the NJL context is
well known and is described in great details in standard references of this
issue (see for example \cite{Klimt-Weise,HK-rev,Bernard} and references
therein). Basically, besides the gap equation (\ref{gap_eq}), we have to solve
a Bethe-Salpeter type equation which in nuclear physics language is known as
Hartree-Fock plus Random Phase Approximation. Such program is of simple
realization when only spin zero mesons are considered (pion and sigma). On the
other hand, the model having vector and axial-vector mesons (rho and $a_{1}$)
implies in additional contributions for the scalar sector. The same will occur
with the vector sector when the tensor coupling is taken into account. As an
example, in the pionic sector, by solving the Bethe-Salpeter equation, we find
a mixing between the pseudoscalar ($\pi$) and the axial-vector ($a_{1}$)
mesons. The pion mass $m_{\pi}$ , is given by solving the condition%
\begin{equation}
\left.  D_{\pi}\left(  p^{2}\right)  \right\vert _{p^{2}=m_{\pi}^{2}%
}=0,\label{m_pi_def}%
\end{equation}
where%
\[
D_{\pi}\left(  p^{2}\right)  =\left[  1-G_{S}\Pi^{PP}\left(  p^{2}\right)
\right]  \left[  1+G_{V}\Pi_{\left(  L\right)  }^{AA}\left(  p^{2}\right)
\right]  +G_{S}G_{V}\left[  \Pi^{AP}\left(  p^{2}\right)  \right]  ^{2}.
\]
Here we have introduced the longitudinal and transverse tensors, defined as
\cite{Klimt-Weise}:
\begin{align*}
\Pi_{\left(  L\right)  }^{AA}\left(  p^{2}\right)   &  =-L^{\mu\nu}\left[
\Pi_{\mu\nu}^{AA}\left(  p^{2}\right)  \right]  ,\\
T^{\mu\nu}\left[  \Pi_{\left(  T\right)  }^{AA}\left(  p^{2}\right)  \right]
&  =-T^{\mu\alpha}T^{\nu\beta}\left[  \Pi_{\alpha\beta}^{AA}\left(
p^{2}\right)  \right]  ,\\
\Pi_{\left(  L\right)  }^{VV}\left(  p^{2}\right)   &  =L^{\mu\nu}\left[
\Pi_{\mu\nu}^{VV}\left(  p^{2}\right)  \right]  ,\\
T^{\mu\nu}\left[  \Pi_{\left(  T\right)  }^{VV}\left(  p^{2}\right)  \right]
&  =T^{\mu\alpha}T^{\nu\beta}\left[  \Pi_{\alpha\beta}^{VV}\left(
p^{2}\right)  \right]  ,\\
\Pi^{AP}\left(  p^{2}\right)   &  =-\frac{p^{\mu}}{p^{2}}\left[  \Pi_{\mu
}^{AP}\left(  p^{2}\right)  \right]  ,
\end{align*}
where%
\[
L^{\mu\nu}=\frac{p^{\mu}p^{\nu}}{p^{2}},\ \ \ T^{\mu\nu}=g^{\mu\nu}%
-\frac{p^{\mu}p^{\nu}}{p^{2}},
\]
are the longitudinal and transverse parts of the polarization functions.

The pion phenomenology is also characterized by the decay constant $f_{\pi}$
and the pion-quark-quark coupling strength $g_{\pi qq}$. Experimentally
$f_{\pi}$ is related to the weak decay $\pi^{\pm}\rightarrow\mu^{\pm}+\nu
_{\mu}$ and is calculated from the vacuum to one-pion axial-vector current
matrix element
\[
\langle0|\overline{\psi}(x)\gamma_{\mu}\gamma_{5}\frac{\tau^{i}}{2}\psi
(x)|\pi^{j}\left(  q\right)  \rangle=i\ f_{\pi}q_{\mu}\delta_{ij}e^{-ipx},
\]
where $|\pi^{j}(q)\rangle$ is a pion state with four-momentum $q$. At one-loop
order, one can express this matrix element in terms of the $\Pi^{AP}$, defined
in Eq.~(\ref{polarization}) by taking $\mathcal{M}_{1}=\gamma_{\mu}\gamma_{5}$
and $\mathcal{M}_{2}=\gamma_{5}$ and the longitudinal part of $\Pi_{\mu\nu
}^{AA}$ ($\mathcal{M}_{1}=\gamma_{\mu}\gamma_{5}$ and $\mathcal{M}_{2}%
=\gamma_{\nu}\gamma_{5}$), as%
\begin{equation}
f_{\pi}=\frac{g_{\pi qq}}{2m_{\pi}}\left[  \Pi^{AP}\left(  m_{\pi}^{2}\right)
\right]  +\frac{\widetilde{g}_{\pi qq}}{4M}\left[  \Pi_{\left(  L\right)
}^{AA}\left(  m_{\pi}^{2}\right)  \right]  ,\label{fpi_def}%
\end{equation}
where the $g_{\pi qq}$ and $\widetilde{g}_{\pi qq}$ are the coupling constants
between pion and quarks in the effective interaction Lagrangian
\cite{Takizawa}%
\[
L_{eff}^{\left(  \pi\right)  }=g_{\pi qq}\overline{\psi}\left(  x\right)
i\gamma_{5}\overrightarrow{\tau}\psi\left(  x\right)  \cdot\overrightarrow
{\pi}\left(  x\right)  +\frac{\widetilde{g}_{\pi qq}}{M}\overline{\psi}\left(
x\right)  \gamma_{\mu}\gamma_{5}\overrightarrow{\tau}\psi\left(  x\right)
\cdot\partial^{\mu}\overrightarrow{\pi}\left(  x\right)  ,
\]
with $\overrightarrow{\pi}\left(  x\right)  $ being the meson field with the
quantum numbers of the pion. The $g_{\pi qq}$ and $\widetilde{g}_{\pi qq}$
coupling constant are related to the residuo of the scattering matrix at the
pion pole ($p^{2}=m_{\pi}^{2}$). The corresponding results are
\begin{align}
g_{\pi qq}^{-2} &  =\left(  -\right)  \frac{\left.  \frac{\partial}{\partial
p^{2}}D_{\pi}\left(  p^{2}\right)  \right\vert _{p^{2}=m_{\pi}^{2}}}%
{G_{S}\left[  1+G_{V}\Pi_{\left(  L\right)  }^{AA}\left(  m_{\pi}^{2}\right)
\right]  },\label{g_pi_1_def}\\
\frac{\widetilde{g}_{\pi qq}}{g_{\pi qq}} &  =\left(  -\right)  \frac
{2MG_{V}\left[  \Pi^{PA}\left(  m_{\pi}^{2}\right)  \right]  }{m_{\pi}\left[
1+G_{V}\Pi_{\left(  L\right)  }^{AA}\left(  m_{\pi}^{2}\right)  \right]
}.\label{g_pi_2_def}%
\end{align}

The vector and axial-vector mesons are described in a very similar way . The
pole of the scattering matrix in the vector channel gives the condition which
determines the rho mass ($m_{\rho}$), i.e.,%
\begin{align}
\left.  D_{\rho}\left(  p^{2}\right)  \right\vert _{p^{2}=m_{\rho}^{2}}  &
=0,\label{m_rho}\\
D_{\rho}\left(  p^{2}\right)   & =1+G_{V}\left[  \Pi_{\left(  T\right)  }%
^{VV}\left(  p^{2}\right)  \right]  .\nonumber
\end{align}
The residuo of the scattering matrix at the rho pole ($p^{2}=m_{\rho}^{2}$) is
related to the coupling constants between rho and quarks in the effective
interaction Lagrangian%
\[
L_{eff}^{\left(  \rho\right)  }=g_{\rho qq}\overline{\psi}\left(  x\right)
\gamma_{\mu}\frac{\tau_{i}}{2}\psi\left(  x\right)  \rho_{i}^{\mu}\left(
x\right)  ,
\]
where $\rho_{i}^{\nu}\left(  x\right)  $ is the vector field. The
rho-quark-quark coupling constant is given by%
\begin{equation}
g_{\rho qq}^{-2}=\frac{1}{4G_{V}}\left.  \frac{\partial}{\partial p^{2}%
}D_{\rho}\left(  p^{2}\right)  \right\vert _{p^{2}=m_{\rho}^{2}}%
.\label{g_rho_q_q}%
\end{equation}
On the other hand, the matrix elements of the electromagnetic current between
vector meson states and the vacuum defines the rho meson decay constant
($f_{\rho}$):%
\[
\left\langle 0\left\vert \overline{\psi}\left(  x\right)  \frac{1}{2}%
\gamma_{\mu}\tau_{3}\psi\left(  x\right)  \right\vert \rho\right\rangle
=\frac{m_{\rho}^{2}}{f_{\rho}}\varepsilon_{\mu},
\]
where $\varepsilon_{\mu}$ is the polarization vector of the $\rho$ meson
field. We have then the following relation%
\begin{equation}
\frac{4m_{\rho}^{2}}{f_{\rho}}=g_{\rho qq}\left.  \left[  \Pi_{\left(
T\right)  }^{VV}\left(  p^{2}\right)  \right]  \right\vert _{p^{2}=m_{\rho
}^{2}}.\label{f_rho}%
\end{equation}
Finally, the simplest effective interaction Lagrangean describing the coupling
of a axial-vector field $a_{1i}^{\mu}\left(  x\right)  $ with quark fields may
be written as%
\[
L_{effe}^{\left(  a_{1}\right)  }=g_{a_{1}qq}\overline{\psi}\left(  x\right)
\gamma_{\mu}\gamma_{5}\tau_{i}\psi\left(  x\right)  a_{1i}^{\mu}\left(
x\right)  .
\]
In the RPA approximation the axial meson ($a_{1}$) mass $m_{a_{1}}$ is given
by the condition:%
\begin{align*}
\left.  D_{a_{1}}\left(  p^{2}\right)  \right\vert _{p^{2}=m_{a_{1}}^{2}}  &
=0,\\
D_{a_{1}}\left(  q^{2}\right)   & =1+G_{V}\left[  \Pi_{\left(  T\right)
}^{AA}\left(  p^{2}\right)  \right]  .
\end{align*}
The $a_{1}$-quark-quark coupling constant is determined through the condition
\[
g_{a_{1}qq}^{-2}=4\left.  \frac{\partial}{\partial p^{2}}\Pi_{\left(
T\right)  }^{AA}\left(  p^{2}\right)  \right\vert _{p^{2}=m_{a_{1}}^{2}}.
\]
In the next section we evaluate the involved amplitudes, within the context of
our strategy. After this we will return to the definitions introduced above to
calculate the corresponding physical quantities.

\section{The evaluation of physical amplitudes}

Let us start with the relevant aspects of the adopted strategy to handle the
divergences, alternative to the traditional regularization techniques. Such
strategy, proposed and developed by O.A. Battistel in Ref. \cite{Orimar-tese},
has a central idea which is to avoid the critical step involved in the
regularization process: the explicit evaluation of divergent integrals. In the
intermediary steps it is assumed the presence of a regularization only in an
implicit way. No specific form for the regularization distribution is adopted
and only very general properties of an acceptable regularization are assumed.
Through the use of an adequate representation for the propagators, the
amplitudes are written in a convenient mathematical form such that when the
integration is taken, all the dependence on the internal lines (arbitrary
momenta) are located in finite integrals and the divergent ones are written in
terms of standard mathematical objects. With this attitude it becomes possible
to identify the properties required for a regularization in order to eliminate
the potentially ambiguous terms as well as those which are potentially
symmetry violating because they appear separated from the finite parts in a
natural way. Following this strategy we search for the properties that a
regularization must have in order to be consistent.

The implementation of the procedure is made by choosing the adequate
representation for the involved propagators. The idea is to write the
propagators in a way that the momentum structure is emphasized just because it
is in the last instance, this structure that is responsible for the behavior
which generates the divergences in the amplitudes. We adopt then an adequate
representation for a propagator carrying momentum $k+k_{i}$ and mass $M$:
\begin{align}
\frac{1}{[(k+k_{i})^{2}-M^{2}]}  & =\sum_{j=0}^{N}\frac{\left(  -1\right)
^{j}\left(  k_{i}^{2}+2k_{i}\cdot k+\lambda^{2}-M^{2}\right)  ^{j}}{\left(
k^{2}-\lambda^{2}\right)  ^{j+1}}\nonumber\\
& +\frac{\left(  -1\right)  ^{N+1}\left(  k_{i}^{2}+2k_{i}\cdot k+\lambda
^{2}-M^{2}\right)  ^{N+1}}{\left(  k^{2}-\lambda^{2}\right)  ^{N+1}\left[
\left(  k+k_{i}\right)  ^{2}-M^{2}\right]  }.\label{fermion_prop}%
\end{align}
Here $k_{i}$ is (in principle) an arbitrary routing momentum of an internal
line in a loop, and $M$ is the fermion mass running in the loop. In the above
identity we have introduced an arbitrary parameter $\lambda$ with dimension of
mass. As it will be shown in the next section, this parameter gives a precise
connection between the divergent and finite parts. It plays the role of a
common scale for the divergent and finite parts of the corresponding Feynman
integrals. The value for $N$ in the equation above must be taken as major or
equal to the highest superficial degree of divergence of the considered theory
or model, if we want to take a unique representation for all involved
propagators. Once this representation is assumed, the integration in the loop
momentum can be introduced (the last Feynman rule). All the Feynman integrals
containing the internal momenta will be present in finite integrals. On the
other hand, the divergent ones will contain only the arbitrary mass scale
$\lambda$ assuming then standard mathematical forms. No divergent integrals
need, in fact, to be solved. Only the tensor reduction must be specified such
that the divergent content of amplitudes will appear as standard irreducible
divergent objects. They need not be calculated since in renormalizable
theories they are completely absorbed in the reparametrization of the theory
and in nonrenormalizable models they will be directly adjusted to
phenomenological parameters in the parametrization of the model in each
specific level of approximation considered. In the calculation process the
regularization plays a secondary role just because it is only necessary to
assume its implicit presence. The tensor reduction of purely divergent
integrals, the unique assumption involving divergent integrals in the
intermediary steps, works like consistency requirements to be imposed over an
eventual regularization distribution. More details about the procedure will be
presented in a moment when the amplitudes are evaluated.

If we wish to follow the procedure mentioned above to evaluate the amplitudes
$T^{S}$, $T^{PP}$,$\ T_{\mu}^{AP}$,$\ T_{\mu\nu}^{VV}$, and $T_{\mu\nu}^{AA}$,
adopting a universal form for the involved propagators of the internal
fermions, we must take the following representation
\begin{align}
S(k+k_{i};\lambda^{2})  & =\frac{\left(  \not k +\not k _{i}\right)
+M}{[(k+k_{i})^{2}-M^{2}]}\nonumber\\
& =\left(  \not k +\not k _{i}+M\right)  \left\{  \frac{1}{\left(
k^{2}-\lambda^{2}\right)  }-\frac{\left(  k_{i}^{2}+2\left(  k_{i}\cdot
k\right)  +\lambda^{2}-M^{2}\right)  }{\left(  k^{2}-\lambda^{2}\right)  ^{2}%
}\right. \nonumber\\
& +\frac{\left(  k_{i}^{2}+2\left(  k_{i}\cdot k\right)  +\lambda^{2}%
-M^{2}\right)  ^{2}}{\left(  k^{2}-\lambda^{2}\right)  ^{3}}-\frac{\left(
k_{i}^{2}+2\left(  k_{i}\cdot k\right)  +\lambda^{2}-M^{2}\right)  ^{3}%
}{\left(  k^{2}-\lambda^{2}\right)  ^{4}}\nonumber\\
& \left.  +\frac{\left(  k_{i}^{2}+2\left(  k_{i}\cdot k\right)  +\lambda
^{2}-M^{2}\right)  ^{4}}{\left(  k^{2}-\lambda^{2}\right)  ^{4}\left[  \left(
k+k_{i}\right)  ^{2}-M^{2}\right]  }\right\}  .\label{fermion_prop2}%
\end{align}
This expression is obtained by taking $N=3$ in the Eq. (\ref{fermion_prop})
and performing the summation over the values of $j$ ($0$, $1$, $2$ and $3$).
Note that the expression above is, in fact, independent of the arbitrary scale
parameter $\lambda$. This can be easily checked by verifying that
\[
\frac{\partial}{\partial\lambda^{2}}S(k+k_{i};\lambda^{2})=0.
\]
Let us now evaluate the amplitudes starting by taking the one corresponding to
the highest divergence degree: the $T^{S}$ defined in (\ref{T_S_def}). First
we construct the quantity $t^{S}$ by performing the Dirac traces and
substituting the expression (\ref{fermion_prop2}) for the propagator. We get
then
\begin{align*}
t^{S}  & =4M\left\{  \frac{1}{\left(  k^{2}-\lambda^{2}\right)  }-2k_{1\alpha
}\left(  \frac{k_{\alpha}}{\left(  k^{2}-\lambda^{2}\right)  ^{2}}\right)
-\left(  \lambda^{2}-M^{2}\right)  \left(  \frac{1}{\left(  k^{2}-\lambda
^{2}\right)  ^{2}}\right)  \right. \\
& \ \ \ \ \ \ \ +k_{1\alpha}k_{1\beta}\left(  \frac{4k_{\alpha}k_{\beta}%
}{\left(  k^{2}-\lambda^{2}\right)  ^{3}}-\frac{g_{\alpha\beta}}{\left(
k^{2}-\lambda^{2}\right)  ^{2}}\right)  +4k_{1\alpha}\left(  k_{1}^{2}%
+\lambda^{2}-M^{2}\right)  \left(  \frac{k_{\alpha}}{\left(  k^{2}-\lambda
^{2}\right)  ^{3}}\right) \\
& \ \ \ \ \ \ \ \left.  +\frac{\left(  k_{1}^{2}+\lambda^{2}-M^{2}\right)
^{2}}{\left(  k^{2}-\lambda^{2}\right)  ^{3}}-\frac{\left(  k_{1}^{2}%
+2k_{1}\cdot k+\lambda^{2}-M^{2}\right)  ^{3}}{\left(  k^{2}-\lambda
^{2}\right)  ^{4}}+\frac{\left(  k_{1}^{2}+2k_{1}\cdot k+\lambda^{2}%
-M^{2}\right)  ^{4}}{\left(  k^{2}-\lambda^{2}\right)  ^{4}\left[  \left(
k+k_{1}\right)  ^{2}-m^{2}\right]  }\right\}  .
\end{align*}
After this, integrating over the loop momenta $k$ we get the amplitude $T^{S}
$ given by
\begin{align}
T^{S}  & =-4M\left\{  i\left[  I_{quad}\left(  \lambda^{2}\right)  \right]
+\left(  M^{2}-\lambda^{2}\right)  i\left[  I_{log}\left(  \lambda^{2}\right)
\right]  \right. \nonumber\\
& \ \ \ \ \ \ \left.  -\frac{1}{16\pi^{2}}\left[  M^{2}-\lambda^{2}-M^{2}%
\ln\left(  \frac{M^{2}}{\lambda^{2}}\right)  \right]  \right\} \nonumber\\
& -ik_{1\alpha}k_{1\beta}\left[  \Delta^{\alpha\beta}\left(  \lambda
^{2}\right)  \right]  ,\label{T_S_1}%
\end{align}
where we have introduced the irreducible divergent objects
\begin{align}
I_{quad}\left(  \lambda^{2}\right)   & =\int\frac{d^{4}k}{\left(  2\pi\right)
^{4}}\frac{1}{\left(  k^{2}-\lambda^{2}\right)  },\label{I_log_def}\\
I_{log}\left(  \lambda^{2}\right)   & =\int\frac{d^{4}k}{\left(  2\pi\right)
^{4}}\frac{1}{\left(  k^{2}-\lambda^{2}\right)  ^{2}},\label{I_quad_def}%
\end{align}
and also the object%
\begin{equation}
\nabla_{\mu\nu}\left(  \lambda^{2}\right)  =\int_{\Lambda}\frac{d^{4}%
k}{\left(  2\pi\right)  ^{4}}\frac{2k_{\nu}k_{\mu}}{\left(  k^{2}%
-M^{2}\right)  ^{2}}-\int_{\Lambda}\frac{d^{4}k}{\left(  2\pi\right)  ^{4}%
}\frac{g_{\mu\nu}}{\left(  k^{2}-M^{2}\right)  }.\label{Delta}%
\end{equation}
To obtain the result (\ref{T_S_1}) we have only integrated the finite
integrals and left out the integrals having odd integrands. Now we evaluate
the two point-functions. First we rewrite the integrand (\ref{integrand}) of
such amplitudes as
\begin{align*}
\pi_{\mathcal{M}_{1}\mathcal{M}_{2}}  & =\frac{1}{D_{12}}\left\{  Tr\left(
\Gamma_{\mathcal{M}_{1}}\gamma_{\alpha}\Gamma_{\mathcal{M}_{2}}\gamma_{\beta
}\right)  k_{\alpha}k_{\beta}\right. \\
& \ \ \ \ \ \ \ \ +\left[  {k}_{2\beta}Tr\left(  \Gamma_{\mathcal{M}_{1}%
}\gamma_{\alpha}\Gamma_{\mathcal{M}_{2}}\gamma_{\beta}\right)  +mTr\left(
\Gamma_{\mathcal{M}_{1}}\gamma_{\alpha}\Gamma_{\mathcal{M}_{2}}\right)
\right]  k_{\alpha}\\
& \ \ \ \ \ \ \ \ +\left[  {k}_{1\alpha}Tr\left(  \Gamma_{\mathcal{M}_{1}%
}\gamma_{\alpha}\Gamma_{\mathcal{M}_{2}}\gamma_{\beta}\right)  +mTr\left(
\Gamma_{\mathcal{M}_{1}}\Gamma_{\mathcal{M}_{2}}\gamma_{\beta}\right)
\right]  k_{\beta}\\
& \ \ \ \ \ \ \ \ +\left[  {k}_{1\alpha}{k}_{2\beta}Tr\left(  \Gamma
_{\mathcal{M}_{1}}\gamma_{\alpha}\Gamma_{\mathcal{M}_{2}}\gamma_{\beta
}\right)  +m{k}_{1\alpha}Tr\left(  \Gamma_{\mathcal{M}_{1}}\gamma_{\alpha
}\Gamma_{\mathcal{M}_{2}}\right)  \right. \\
& \ \ \ \ \ \ \left.  \left.  +m{k}_{2\beta}Tr\left(  \Gamma_{\mathcal{M}_{1}%
}\Gamma_{\mathcal{M}_{2}}\gamma_{\beta}\right)  +m^{2}Tr\left(  \Gamma
_{\mathcal{M}_{1}}\Gamma_{\mathcal{M}_{2}}\right)  \right]  \right\}  ,
\end{align*}
where we have introduced the notation $D_{12}=\left[  \left(  k+k_{1}\right)
^{2}-m^{2}\right]  \left[  \left(  k+k_{2}\right)  ^{2}-m^{2}\right]  $. Now,
by using the representation (\ref{fermion_prop2}) for the fermion-propagators,
performing the Dirac traces with the appropriate $\Gamma_{\mathcal{M}_{1}}$
and $\Gamma_{\mathcal{M}_{2}}$ chosen, integrating over the loop momentum $k $
and, again dropping out the odd integrals, the two point-functions which are
necessary in this work are given by
\begin{equation}
T_{\mu}^{PA}=-4Mq_{\mu}\left\{  i\left[  I_{log}\left(  \lambda^{2}\right)
\right]  +\frac{1}{16\pi^{2}}\left[  Z_{0}\left(  q^{2},M^{2};\lambda
^{2}\right)  \right]  \right\}  ,\label{T_AP_1}%
\end{equation}%
\begin{align}
T^{PP} &  =4i\left[  I_{quad}\left(  \lambda^{2}\right)  \right]  -4\left(
\lambda^{2}-M^{2}\right)  i\left[  I_{log}\left(  \lambda^{2}\right)  \right]
\nonumber\\
&  +\frac{1}{4\pi^{2}}\left[  \lambda^{2}-M^{2}-M^{2}\ln\left(  \frac
{\lambda^{2}}{M^{2}}\right)  \right] \nonumber\\
&  -2q^{2}\left\{  i\left[  I_{log}\left(  \lambda^{2}\right)  \right]
+\frac{1}{16\pi^{2}}\left[  Z_{0}\left(  q^{2},M^{2};\lambda^{2}\right)
\right]  \right\} \nonumber\\
&  +i\left(  q_{\alpha}q_{\beta}+Q_{\alpha}Q_{\beta}\right)  \left[
\Delta^{\alpha\beta}\left(  \lambda^{2}\right)  \right]  ,\label{T_PP_1}%
\end{align}%
\begin{align}
T_{\mu\nu}^{VV}  & =\frac{4}{3}\left(  g_{\mu\nu}q^{2}-q_{\mu}q_{\nu}\right)
i\left[  I_{log}\left(  \lambda^{2}\right)  \right] \nonumber\\
& -\frac{\left(  g_{\mu\nu}q^{2}-q_{\mu}q_{\nu}\right)  }{2\pi^{2}}\left[
Z_{2}\left(  q^{2},M^{2};\lambda^{2}\right)  -Z_{1}\left(  q^{2},M^{2}%
;\lambda^{2}\right)  \right]  +A_{\mu\nu}\left(  \lambda^{2}\right)
,\label{T_VV_1}%
\end{align}%
\begin{align}
T_{\mu\nu}^{AA}  & =\frac{4}{3}\left(  g_{\mu\nu}q^{2}-q_{\mu}q_{\nu}\right)
i\left[  I_{log}\left(  \lambda^{2}\right)  \right] \nonumber\\
& -\frac{\left(  g_{\mu\nu}q^{2}-q_{\mu}q_{\nu}\right)  }{2\pi^{2}}\left[
Z_{2}\left(  q^{2},M^{2};\lambda^{2}\right)  -Z_{1}\left(  q^{2},M^{2}%
;\lambda^{2}\right)  \right] \nonumber\\
& -g_{\mu\nu}M^{2}\left\{  8i\left[  I_{log}\left(  \lambda^{2}\right)
\right]  +\frac{1}{2\pi^{2}}\left[  Z_{0}\left(  q^{2},M^{2};\lambda
^{2}\right)  \right]  \right\}  +A_{\mu\nu}\left(  \lambda^{2}\right)
.\label{T_AA_1}%
\end{align}
In the above expressions we have introduced the $Z_{k}\left(  q^{2}%
,M^{2};\lambda^{2}\right)  $ finite structures
\begin{equation}
Z_{k}\left(  q^{2},M^{2};\lambda^{2}\right)  =\int_{0}^{1}dz\,z^{k}%
\log\left\{  \frac{q^{2}z(1-z)-M^{2}}{-\lambda^{2}}\right\}  .\label{Z_k}%
\end{equation}
The quantity $A_{\mu\nu}$ represents the expression
\begin{align}
A_{\mu\nu}  & =4\left[  \nabla_{\mu\nu}\left(  \lambda^{2}\right)  \right]
+q^{\alpha}q^{\beta}\left[  \frac{1}{3}\square_{\alpha\beta\mu\nu}\left(
\lambda^{2}\right)  +\frac{1}{3}g_{\alpha\nu}\Delta_{\mu\beta}\left(
\lambda^{2}\right)  +g_{\alpha\mu}\Delta_{\beta\nu}\left(  \lambda^{2}\right)
\right. \nonumber\\
& \left.  -g_{\mu\nu}\Delta_{\alpha\beta}\left(  \lambda^{2}\right)  -\frac
{2}{3}g_{\alpha\beta}\Delta_{\mu\nu}\left(  \lambda^{2}\right)  \right]
\nonumber\\
& +\left(  q^{\alpha}Q^{\beta}-Q^{\alpha}q^{\beta}\right)  \left[  \frac{1}%
{3}\square_{\alpha\beta\mu\nu}\left(  \lambda^{2}\right)  +\frac{1}{3}%
g_{\nu\alpha}\Delta_{\mu\beta}\left(  \lambda^{2}\right)  +\frac{1}%
{3}g_{\alpha\mu}\Delta_{\beta\nu}\left(  \lambda^{2}\right)  \right]
\nonumber\\
& +Q^{\alpha}Q^{\beta}\left[  \square_{\alpha\beta\mu\nu}\left(  \lambda
^{2}\right)  -g_{\mu\beta}\Delta_{\nu\alpha}\left(  \lambda^{2}\right)
-g_{\alpha\mu}\Delta_{\beta\nu}\left(  \lambda^{2}\right)  -3g_{\mu\nu}%
\Delta_{\alpha\beta}\left(  \lambda^{2}\right)  \right]  .
\end{align}
where we have introduced two new divergent structures defined as%
\begin{align}
\square_{\alpha\beta\mu\nu}\left(  \lambda^{2}\right)   & =\int_{\Lambda}%
\frac{d^{4}k}{\left(  2\pi\right)  ^{4}}\frac{24k_{\mu}k_{\nu}k_{\alpha
}k_{\beta}}{\left(  k^{2}-M^{2}\right)  ^{4}}-g_{\alpha\beta}\int_{\Lambda
}\frac{d^{4}k}{\left(  2\pi\right)  ^{4}}\frac{4k_{\mu}k_{\nu}}{\left(
k^{2}-M^{2}\right)  ^{3}}\nonumber\\
& -g_{\alpha\nu}\int_{\Lambda}\frac{d^{4}k}{\left(  2\pi\right)  ^{4}}%
\frac{4k_{\beta}k_{\mu}}{\left(  k^{2}-M^{2}\right)  ^{3}}-g_{\alpha\mu}%
\int_{\Lambda}\frac{d^{4}k}{\left(  2\pi\right)  ^{4}}\frac{4k_{\beta}k_{\nu}%
}{\left(  k^{2}-M^{2}\right)  ^{3}},\\
\Delta_{\mu\nu}\left(  \lambda^{2}\right)   & =\int_{\Lambda}\frac{d^{4}%
k}{\left(  2\pi\right)  ^{4}}\frac{4k_{\mu}k_{\nu}}{\left(  k^{2}%
-M^{2}\right)  ^{3}}-\int_{\Lambda}\frac{d^{4}k}{\left(  2\pi\right)  ^{4}%
}\frac{g_{\mu\nu}}{\left(  k^{2}-M^{2}\right)  ^{2}}.
\end{align}

Note that when the integration sign was introduced (the last Feynman rule) we
performed the indicated operation only in the finite integrals. On the other
hand, for those structures which have emerged divergent, no additional
operation is performed. This means that in the expressions for the calculated
amplitudes five quantities remain undefined. They are: $I_{quad}\left(
\lambda^{2}\right)  $, $I_{log}\left(  \lambda^{2}\right)  $, $\Delta
_{\alpha\beta}\left(  \lambda^{2}\right)  $, $\nabla_{\alpha\beta}\left(
\lambda^{2}\right)  $, and $\square_{\alpha\beta\mu\nu}\left(  \lambda
^{2}\right)  $. The objects $\Delta_{\alpha\beta}\left(  \lambda^{2}\right)
$, $\nabla_{\alpha\beta}\left(  \lambda^{2}\right)  $, and $\square
_{\alpha\beta\mu\nu}\left(  \lambda^{2}\right)  $ are differences between
integrals having the same degree of divergence. In principle, to obtain a
value for these objects, the integrand must be made finite through the
assumption of a regularization distribution. This process can be schematically
represented as
\begin{equation}
\int\frac{d^{4}k}{\left(  2\pi\right)  ^{4}}f(k)\rightarrow\int\frac{d^{4}%
k}{\left(  2\pi\right)  ^{4}}f(k)\left\{  \lim_{\Lambda_{i}^{2}\rightarrow
\infty}G_{\Lambda_{i}}\left(  k,\Lambda_{i}^{2}\right)  \right\}
=\int_{\Lambda}\frac{d^{4}k}{\left(  2\pi\right)  ^{4}}f(k).
\end{equation}
For such regularization distribution to be acceptable it must be even in the
integrating momentum $k$ in order to be consistent with the Lorentz invariance
maintenance. This is the reason why we have excluded divergent integrals
having odd integrands when the integration sign was introduced. The
regularization distribution must also possess the well defined limit
\begin{equation}
\lim_{\Lambda_{i}^{2}\rightarrow\infty}G_{\Lambda_{i}}\left(  k^{2}%
,\Lambda_{i}^{2}\right)  =1,
\end{equation}
which allows us to connect with the original integral. Here $\Lambda_{i}^{2} $
are parameters characterizing the regularization distribution. In what follows
we will note that the evaluation of divergent integrals is in fact not really necessary.

Following our reasoning line we now search for additional properties that a
consistent regularization must have in order to be consistent. First we note
that there are potentially ambiguous terms in the calculated amplitudes. They
are present in the amplitudes $T^{S}$,\ $T^{PP}$, $T_{\mu\nu}^{VV}$, and
$T_{\mu\nu}^{AA}$ and are given by
\begin{align*}
\left(  T^{S}\right)  _{ambi}  & =k_{1\alpha}k_{1\beta}\left[  \Delta
^{\alpha\beta}\left(  \lambda^{2}\right)  \right]  ,\\
\left(  T^{PP}\right)  _{ambi}  & =-Q_{\alpha}Q_{\beta}\left[  \Delta
^{\alpha\beta}\left(  \lambda^{2}\right)  \right]  ,\\
\left(  T_{\mu\nu}^{VV}\right)  _{ambi}  & =\left(  T_{\mu\nu}^{AA}\right)
_{ambi}=\left(  q^{\alpha}Q^{\beta}-Q^{\alpha}q^{\beta}\right) \\
& \times\left[  \frac{1}{3}\square_{\alpha\beta\mu\nu}\left(  \lambda
^{2}\right)  +\frac{1}{3}g_{\nu\alpha}\Delta_{\mu\beta}\left(  \lambda
^{2}\right)  +\frac{1}{3}g_{\alpha\mu}\Delta_{\beta\nu}\left(  \lambda
^{2}\right)  \right] \\
& +Q^{\alpha}Q^{\beta}\left[  \square_{\alpha\beta\mu\nu}\left(  \lambda
^{2}\right)  -g_{\mu\beta}\Delta_{\nu\alpha}\left(  \lambda^{2}\right)
-g_{\alpha\mu}\Delta_{\beta\nu}\left(  \lambda^{2}\right)  -3g_{\mu\nu}%
\Delta_{\alpha\beta}\left(  \lambda^{2}\right)  \right]  .
\end{align*}
These terms are ambiguous due two reasons. First, the dependence on the
momentum $Q=k_{1}+k_{2}$, in the two-point functions, or $k_{1}$ in the
one-point function, implies ambiguity because this quantity is completely
undefined and dependent on the choices for the internal momenta routing. For
the second, the quantity $\Delta^{\alpha\beta}\left(  \lambda^{2}\right)  $
may only be dependent on the arbitrary scale parameter $\lambda$ which is also
a choice. On the other hand, there are terms in the $T^{PP}$, $T_{\mu\nu}%
^{VV}$, and $T_{\mu\nu}^{AA}$ amplitudes which are nonambiguous concerning the
internal momenta choices such as $q_{\alpha}q_{\beta}\left[  \Delta
^{\alpha\beta}\left(  \lambda^{2}\right)  \right]  $ and $\nabla_{\mu\nu
}\left(  \lambda^{2}\right)  $. This means that, in order to eliminate all the
ambiguous terms, we have no option rather than require the following property
for a consistent regularization (for an extensive discussion see the Ref.
\cite{Orimar-PRD-NJL})
\begin{equation}
\Delta_{reg}^{\alpha\beta}\left(  \lambda^{2}\right)  =\nabla_{reg}%
^{\alpha\beta}\left(  \lambda^{2}\right)  =\square_{reg}^{\alpha\beta\mu\nu
}\left(  \lambda^{2}\right)  =0,\label{CR}%
\end{equation}
which we denominate consistency relations (CR). These properties are an
unavoidable requirement if we want to take an additional step in the
evaluation of the considered physical amplitudes. In fact, these properties
are satisfied within the context of DR and Pauli-Villars methods (see Ref.
\cite{Orimar-PRD1}), for example. We can understand these constraints as
follows: if we cannot find such regularization distribution, the perturbative
evaluation of physical amplitudes does not make any sense since the results
will be dependent on intermediary arbitrary choices. Predictions cannot be
made since undefined quantities must be chosen before the description of a
certain phenomenology. The analysis of symmetry relations reveals that the
same conditions are required to avoid symmetry violations.

After this brief discussion we can define the consistently regularized
amplitudes by adopting the conditions (\ref{CR}):
\begin{align}
\mathcal{T}^{S}  & =-4M\left\{  i\left[  I_{quad}\left(  \lambda^{2}\right)
\right]  +\left(  M^{2}-\lambda^{2}\right)  i\left[  I_{log}\left(
\lambda^{2}\right)  \right]  \right. \nonumber\\
& \ \ \ \ \ \ \left.  -\frac{1}{16\pi^{2}}\left[  M^{2}-\lambda^{2}-M^{2}%
\ln\left(  \frac{M^{2}}{\lambda^{2}}\right)  \right]  \right\}  ,\label{T_S_2}%
\end{align}%
\begin{equation}
\mathcal{T}_{\mu}^{AP}=-4Mq_{\mu}\left\{  i\left[  I_{log}\left(  \lambda
^{2}\right)  \right]  +\frac{1}{16\pi^{2}}\left[  Z_{0}\left(  q^{2}%
,M^{2};\lambda^{2}\right)  \right]  \right\}  ,\label{T_AP_2}%
\end{equation}%
\begin{align}
\mathcal{T}^{PP} &  =4i\left[  I_{quad}\left(  \lambda^{2}\right)  \right]
-4\left(  \lambda^{2}-M^{2}\right)  i\left[  I_{log}\left(  \lambda
^{2}\right)  \right] \nonumber\\
&  +\frac{1}{4\pi^{2}}\left[  \lambda^{2}-M^{2}-M^{2}\ln\left(  \frac
{\lambda^{2}}{M^{2}}\right)  \right] \nonumber\\
&  -2q^{2}\left\{  i\left[  I_{log}\left(  \lambda^{2}\right)  \right]
+\frac{1}{16\pi^{2}}\left[  Z_{0}\left(  q^{2},M^{2};\lambda^{2}\right)
\right]  \right\}  ,\label{T_PP_2}%
\end{align}%
\begin{align}
\mathcal{T}_{\mu\nu}^{VV}  & =\frac{4}{3}\left(  g_{\mu\nu}q^{2}-q_{\mu}%
q_{\nu}\right)  i\left[  I_{log}\left(  \lambda^{2}\right)  \right]
\nonumber\\
& -\frac{\left(  g_{\mu\nu}q^{2}-q_{\mu}q_{\nu}\right)  }{2\pi^{2}}\left[
Z_{2}\left(  q^{2},M^{2};\lambda^{2}\right)  -Z_{1}\left(  q^{2},M^{2}%
;\lambda^{2}\right)  \right]  ,\label{T_VV_2}%
\end{align}
and%
\begin{align}
\mathcal{T}_{\mu\nu}^{AA}  & =\frac{4}{3}\left(  g_{\mu\nu}q^{2}-q_{\mu}%
q_{\nu}\right)  i\left[  I_{log}\left(  \lambda^{2}\right)  \right]
\nonumber\\
& -\frac{\left(  g_{\mu\nu}q^{2}-q_{\mu}q_{\nu}\right)  }{2\pi^{2}}\left[
Z_{2}\left(  q^{2},M^{2};\lambda^{2}\right)  -Z_{1}\left(  q^{2},M^{2}%
;\lambda^{2}\right)  \right] \nonumber\\
& -g_{\mu\nu}M^{2}\left\{  8i\left[  I_{log}\left(  \lambda^{2}\right)
\right]  +\frac{1}{2\pi^{2}}\left[  Z_{0}\left(  q^{2},M^{2};\lambda
^{2}\right)  \right]  \right\}  .\label{T_AA_2}%
\end{align}
Even if the property (\ref{CR}) is fulfilled, it remains yet the possibility
of the amplitudes being dependent on the choice for the arbitrary scale
parameter $\lambda$ when the quantities $I_{quad}\left(  \lambda^{2}\right)  $
and $I_{\log}\left(  \lambda^{2}\right)  $ are evaluated within the context of
a certain regularization. Then we can ask ourselves about the additional
conditions to be satisfied by a regularization in order to produce scale
independent amplitudes. For this purpose we take initially the $\mathcal{T}%
_{\mu}^{AP}$ amplitude. Obviously, scale independence must be required for the
full amplitude (divergent and finite parts), i.e.,
\[
\frac{\partial\mathcal{T}_{\mu}^{AP}}{\partial\lambda^{2}}=0.
\]
The derivative of the finite part can be performed such that the above
imposition states the property
\begin{equation}
\frac{\partial}{\partial\lambda^{2}}\left[  I_{\log}\left(  \lambda
^{2}\right)  \right]  =-\frac{i}{16\pi^{2}\lambda^{2}}.\label{rel_1}%
\end{equation}
The argument of scale independence can be applied to the $\mathcal{T}^{S}$
amplitude resulting in a property relating in a precise way the irreducible
divergent objects
\begin{equation}
\frac{\partial}{\partial\lambda^{2}}\left[  I_{quad}\left(  \lambda
^{2}\right)  \right]  =I_{\log}\left(  \lambda^{2}\right)  .\label{rel_2}%
\end{equation}
At this point we can ask ourselves the following: what do the requirements
mean? (\ref{rel_1}) and (\ref{rel_2})? They represent two additional
consistency requirements to be imposed over an eventual regularization
distribution in order to classify it as a consistent regularization. These
conditions can be viewed in a very clear way when we relate the irreducible
divergent object in two different mass scales $\lambda$ and $\lambda_{0}$
\begin{align}
i\left[  I_{\log}\left(  \lambda^{2}\right)  \right]   & =i\left[
I_{log}\left(  \lambda_{0}^{2}\right)  \right]  -\frac{1}{16\pi^{2}}\ln\left(
\frac{\lambda_{0}^{2}}{\lambda^{2}}\right)  ,\label{scale_1}\\
i\left[  I_{quad}\left(  \lambda^{2}\right)  \right]   & =i\left[
I_{quad}\left(  \lambda_{0}^{2}\right)  \right]  +\left(  \lambda^{2}%
-\lambda_{0}^{2}\right)  i\left[  I_{log}\left(  \lambda_{0}^{2}\right)
\right] \nonumber\\
& -\frac{1}{16\pi^{2}}\left[  \lambda^{2}-\lambda_{0}^{2}+\lambda^{2}%
\ln\left(  \frac{\lambda_{0}^{2}}{\lambda^{2}}\right)  \right]
.\label{scale_2}%
\end{align}
These relations can be obtained from the $\mathcal{T}^{S}$ and $\mathcal{T}%
_{\mu}^{AP}$ amplitudes, Eqs. (\ref{T_S_2}) and (\ref{T_PP_2}), by first
evaluating $\mathcal{T}^{S}$ at $k_{1}=0$ and $\mathcal{T}^{PP}$ at $q^{2}=0$.
Note that the imposition of independence with $\lambda$ leads us in a natural
way to the properties (\ref{rel_1}) and (\ref{rel_2}). Once the above equation
allows us to relate the irreducible divergent object in different scales we
denominate them as scale relations. The violation of these properties will
result in the breaking of scale properties of physical amplitudes which
possess the same status of symmetry violations.

The scale relations allow us to see a very important consequence of the
consistency requirements for the irreducible divergent objects. If we separate
the parts relative to the dependence on the two different mass scale
parameters into two different sides of the equation, we can easily conclude
that the equality of both sides can be only justified if both sides are
simultaneously equal to the same constant, which means that
\begin{align}
i\left[  I_{\log}\left(  \lambda^{2}\right)  \right]  -\frac{1}{16\pi^{2}}%
\ln\left(  \lambda^{2}\right)   & \equiv C_{1},\label{C1}\\
i\left[  I_{quad}\left(  \lambda^{2}\right)  \right]  -\lambda^{2}i\left[
I_{log}\left(  \lambda^{2}\right)  \right]  +\frac{1}{16\pi^{2}}\lambda^{2}  &
\equiv C_{2},\label{rel_3}%
\end{align}
where $C_{1}$ and $C_{2}$ are arbitrary constants. These two equations will be
useful in the next Section where we study the implications of the scale
properties to the formulation of the NJL model within our approach.

\section{Phenomenological predictions}

After the developments performed in the preceding sections we are at the
position of considering the parametrization of the NJL model, at the
considered level of approximation, and after this appreciating the
corresponding phenomenological predictions. These aspects must be considered
by studying the consequences of the consistency requirements stated in last section.

Let us start by considering the simplest case, the chirally symmetric one
($m_{0}=0$). As we have discussed above, due to the nonrenormalizable
character of the NJL model, the remaining undefined objects, $I_{quad}\left(
\lambda^{2}\right)  $ and $I_{log}\left(  \lambda^{2}\right)  $, must be
related to the physical parameters which are taken as inputs of the model. In
the present work, as it is usual, we take the quark condensate $\langle
\overline{\psi}\psi\rangle$, the pion decay $f_{\pi}$, and the vector coupling
$G_{V}$ as being the phenomenological inputs of the model. In this line of
reasoning, it is easy to see that $I_{quad}\left(  M^{2}\right)  $ can be
directly related to the quark condensate. This fact can be noted from Eqs.
(\ref{cond}) and (\ref{T_S_def}). The result is%
\begin{equation}
i\left[  I_{quad}\left(  M^{2}\right)  \right]  =-\frac{\left\langle
\overline{\psi}\psi\right\rangle }{4N_{c}M}.\label{I_quad}%
\end{equation}
On the other hand, the parametrization of the $I_{\log}\left(  M^{2}\right)  $
in terms of the inputs parameters is not so obvious and immediate. Due to this
let us work out this parametrization in some detail. At first we note the
relation%
\[
\Pi^{PA}\left(  p^{2}\right)  =-\frac{\sqrt{p^{2}}}{2M}\left[  \Pi_{\left(
L\right)  }^{AA}\left(  p^{2}\right)  \right]  ,
\]
which, by using Eqs. (\ref{fpi_def}) and (\ref{g_pi_2_def}), allows to write%
\begin{align}
f_{\pi}  & =-\frac{g_{\pi qq}}{4M}\left(  1-\frac{\widetilde{g}_{\pi qq}%
}{g_{\pi qq}}\right)  \left[  \Pi_{\left(  L\right)  }^{AA}\left(  0\right)
\right]  ,\label{fpi_1}\\
\frac{\widetilde{g}_{\pi qq}}{g_{\pi qq}}  & =\frac{G_{V}\left[  \Pi_{\left(
L\right)  }^{AA}\left(  0\right)  \right]  }{\left[  1+G_{V}\Pi_{\left(
L\right)  }^{AA}\left(  0\right)  \right]  },\label{g_pi_1_2}%
\end{align}
remembering that in the chirally symmetric case $p^{2}=m_{\pi}^{2}=0$. In the
same way we may write%
\begin{align*}
D_{\pi}\left(  p^{2}\right)   & =-\frac{G_{S}}{4M^{2}}p^{2}\left[
\Pi_{\left(  L\right)  }^{AA}\left(  p^{2}\right)  \right]  ,\\
\frac{\partial}{\partial p^{2}}D_{\pi}\left(  p^{2}\right)   & =-\frac{G_{S}%
}{4M^{2}}\left[  \Pi_{\left(  L\right)  }^{AA}\left(  p^{2}\right)  \right]
+\frac{N_{c}}{4\pi^{2}}G_{S}p^{2}\left[  Y_{1}\left(  p^{2},M^{2}\right)
\right]  ,
\end{align*}
where we have used
\begin{align*}
\Pi^{PP}\left(  p^{2}\right)   & =8N_{c}i\left[  I_{quad}\left(  M^{2}\right)
\right]  +\frac{p^{2}}{4M^{2}}\left[  \Pi_{\left(  L\right)  }^{AA}\left(
p^{2}\right)  \right]  ,\\
Y_{1}\left(  p^{2},M^{2}\right)   & =\frac{\partial}{\partial p^{2}}%
Z_{0}\left(  p^{2},M^{2};M^{2}\right)  .
\end{align*}
The expressions above and the Eq. (\ref{g_pi_1_def}) lead us to write%
\[
g_{\pi qq}^{-2}=\frac{\left[  \Pi_{\left(  L\right)  }^{AA}\left(  0\right)
\right]  }{4M^{2}\left[  1+G_{V}\Pi_{\left(  L\right)  }^{AA}\left(  0\right)
\right]  }.
\]
From Eqs. (\ref{fpi_1}), (\ref{g_pi_1_2}) we get%
\[
\Pi_{\left(  L\right)  }^{AA}\left(  0\right)  =\frac{4f_{\pi}^{2}}%
{1-4G_{V}f_{\pi}^{2}}.
\]
Since%
\[
\Pi_{\left(  L\right)  }^{AA}\left(  0\right)  =-16N_{c}M^{2}i\left[  I_{\log
}\left(  M^{2}\right)  \right]  ,
\]
we finally arrived at the searched parametrization for $I_{\log}\left(
M^{2}\right)  .$ It is given by
\begin{equation}
i\left[  I_{\log}\left(  M^{2}\right)  \right]  =-\frac{1}{4N_{c}M^{2}}%
\frac{f_{\pi}^{2}}{1-4G_{V}f_{\pi}^{2}}.\label{I_log}%
\end{equation}
It is interesting to note the following simple (and exact) relations involving
the effective pion coupling constant%
\begin{align*}
M  & =g_{\pi qq}f_{\pi},\\
\widetilde{g}_{\pi qq}  & =4G_{V}Mf_{\pi}.
\end{align*}
So, the Eqs. (\ref{I_quad}) and (\ref{I_log}) state the required relations
between the two remaining undefined quantities and the input parameters.

Next we consider a very important aspect of our formulation. In the last
section we have shown that the objects $I_{quad}\left(  M^{2}\right)  $ and
$I_{log}\left(  M^{2}\right)  $ are not independent and precise relations
between them was stated. Having this in mind we now, by using the results
(\ref{I_quad}) and (\ref{I_log}) in Eq.(\ref{rel_3}), obtain a simple
algebraic equation for the quark mass:
\begin{equation}
M^{3}+\alpha M+\beta=0,\label{algebraic}%
\end{equation}
where we have defined
\[
\alpha=\frac{4\pi^{2}}{3}\frac{f_{\pi}^{2}}{1-4G_{V}f_{\pi}^{2}}%
-C,\ \ \ \ \ \beta=-\frac{4\pi^{2}}{3}\left\langle \overline{\psi}%
\psi\right\rangle ,\ \ \ C=16\pi^{2}C_{2}.
\]
The cubic algebraic equation (\ref{algebraic}) gives all the possible values
for the constituent quark mass $M$ - given the values for the coefficients
$\alpha$ and $\beta$ - which are consistent with the scale independence
requirements. We note that both the sign and the value of $\alpha$ are
dependent on the value assumed by the constant $C$. Therefore, at this point
one could conclude that the physical implications of the model are dependent
on the choice of an arbitrary constant since the solutions of the
Eq.~(\ref{algebraic}) is obviously sensitive to changes of the coefficient
$\alpha$. Putting that in different words, at this stage, it seems that the
physical implications are definitely regularization dependent since the
parameter $C$ is determined by the specific form of the regularization.
Remembering that the regularizations considered as acceptable at this stage
are only those that satisfy the Consistency Relations, which eliminates the
ambiguous and symmetry violating terms,\ as well as obey the properties for
the irreducible divergent objects, which guaranty the scale independence.
However, a more careful analysis must be made since the equation above is a
polynomial of the third degree and, consequently, a very large lack of
possibility for the roots exists depending on the values for the coefficients
of the different powers of $M$. So, the conclusion stated above is premature
and may be wrong.

In order to see what the real situation is, we note that the
Eq.~(\ref{algebraic}), for $\alpha$ and $\beta$ real, has three solutions
which we call $M_{1}$, $M_{2}$ and $M_{3}$. They are:
\begin{align}
M_{1}  & =S+T,\label{sol_1}\\
M_{2}  & =-\frac{1}{2}\left(  S+T\right)  +i\frac{\sqrt{3}}{2}\left(
S-T\right)  ,\label{sol_2}\\
M_{3}  & =-\frac{1}{2}\left(  S+T\right)  -i\frac{\sqrt{3}}{2}\left(
S-T\right)  ,\label{sol_3}%
\end{align}
where
\[
S=\sqrt[3]{-\frac{\beta}{2}+\sqrt{\Delta}},\ \ \ \ \ T=\sqrt[3]{-\frac{\beta
}{2}-\sqrt{\Delta}},\ \ \ \ \ \Delta=\frac{\alpha^{3}}{27}+\frac{\beta^{2}}%
{4}.
\]
There are then three cases, depending on the value assumed by $\Delta$, which
we now study in details. Firstly, for $\Delta>0$ we see that there are no real
positive solutions, thus no physical solution exists if we recognize that only
positive values of $M$ make sense. Secondly, for $\Delta<0$ we have two real
positive roots, $M_{1}$ and $M_{3}$, which can be written as
\begin{align}
M_{1}  & =A\cos\left(  \frac{\theta}{3}\right)  ,\label{M1}\\
M_{3}  & =A\cos\left(  \frac{\theta}{3}+\frac{4\pi}{3}\right)  ,\label{M2}%
\end{align}
where
\begin{align*}
A  & =2\sqrt{\frac{C}{3}-\frac{4\pi^{2}}{9}\frac{f_{\pi}^{2}}{1-4G_{V}f_{\pi
}^{2}}},\\
\cos\theta & =\frac{6\pi^{2}\left\langle \overline{\psi}\psi\right\rangle
}{\sqrt{3\left(  C-\frac{4\pi^{2}}{3}\frac{f_{\pi}^{2}}{1-4G_{V}f_{\pi}^{2}%
}\right)  ^{3}}}.
\end{align*}
This is not a desirable situation just because, for this case, the same set of
parameters imply two different values for the dynamically generated quark
mass. This is unacceptable from the physical point of view. Finally, for
$\Delta=0$ there is just one real positive root. This is a very attractive
possibility. It remains to verify if the values for the quantities involved in
the cancellation of $\Delta$ are reasonable ones. Looking at the equation for
$\Delta$ we note that, in order to achieve this situation, the value for the
arbitrary parameter must be fixed. In this sense we can say that, in order to
obtain a unique solution for the quark mass, we have to pay the price of
fixing the value for the arbitrary parameter.

Now we can invert the interpretation. There is an arbitrary parameter and we
consider the total range of values for it. We note then that for values which
are minor than the critical one given (exactly) by%
\begin{equation}
C_{crit}=\frac{4\pi^{2}}{3}\frac{f_{\pi}^{2}}{1-4G_{V}f_{\pi}^{2}}%
+\sqrt[3]{12\pi^{4}\left\langle \overline{\psi}\psi\right\rangle ^{2}}.
\end{equation}
we have no solutions for the constituent quark mass. For values which are
major than the one above we have two real values for the mass corresponding to
the same set of parameters and, only $C_{crit}$ will lead us to a unique value
for the mass, which is given by
\begin{equation}
M_{crit}=\sqrt[3]{-\frac{2\pi^{2}}{N_{c}}\left\langle \overline{\psi}%
\psi\right\rangle }.\label{M_c}%
\end{equation}
At this point it seems the question of choosing the adequate solution makes no
sense. The answer is certainly obvious: the critical condition naturally fixes
the constant $C$, the last arbitrary parameter still remaining in the model,
and determines the value of the constituent quark mass. Within this point of
view the model becomes predictive in a sense that all the arbitrariness,
involved in the manipulations of the divergent integrals, have disappeared due
to the consistency relations, the scale properties of the irreducible
divergent objects and by the existence of a critical condition.

Now we can obtain the remaining parameter of the model predictions in chirally
symmetric case: the coupling $G_{S}$. It is given by
\begin{equation}
G_{S}=\frac{1}{2}\sqrt[3]{\frac{2\pi^{2}}{3\left\langle \overline{\psi}%
\psi\right\rangle ^{2}}}.
\end{equation}
We note that in our approach the constituent quark mass $M$ and the coupling
$G_{S}$ values depend on the quark condensate $\left\langle \overline{\psi
}\psi\right\rangle $ only. This is a very attractive a new result and
constitutes a relevant difference between our approach and the traditional
ones based on cut-off regularizations. If we assume $\left\langle
\overline{\psi}\psi\right\rangle =\left(  -250.0\;\mathrm{MeV}\right)  ^{3}$
and $f_{\pi}=93.0\;\mathrm{MeV}$ as the two first inputs we obtain
$M_{crit}\simeq468.4\;\mathrm{MeV}$ and $G_{S}\simeq15.0\;\mathrm{GeV}^{-2}$
and $g_{\pi qq}\simeq5.0$. These values are in good accordance with the
expected ones.

Let us now turn our attention to the determination of the vector parameters.
For this purpose we first consider the vector-vector polarization function
evaluated at $p^{2}=m_{\rho}^{2}$. The result is%
\begin{align*}
\Pi_{\left(  T\right)  }^{VV}\left(  m_{\rho}^{2}\right)   &  =\frac{1}%
{6\pi^{2}}m_{\rho}^{2}+\frac{1}{2\pi^{2}}\left(  m_{\rho}^{2}+2M^{2}\right)
\left[  Z_{0}\left(  m_{\rho}^{2},M^{2};M^{2}\right)  \right] \\
&  -\frac{m_{\rho}^{2}}{6M^{2}}\left[  \Pi_{\left(  L\right)  }^{AA}\left(
0\right)  \right]  ,
\end{align*}
which substituted in Eq. (\ref{m_rho}) furnish a transcendental equation which
determines the $\rho^{0}$ mass%
\[
m_{\rho}^{2}=-\frac{2\pi^{2}}{G_{V}}\left\{  \frac{1}{3}+\left(
1+\frac{2M^{2}}{m_{\rho}^{2}}\right)  \left[  Z_{0}\left(  m_{\rho}^{2}%
,M^{2};M^{2}\right)  \right]  -\frac{\pi^{2}}{3M^{2}}\left[  \Pi_{\left(
L\right)  }^{AA}\left(  0\right)  \right]  \right\}  ^{-1}.
\]
In the same way from Eqs. (\ref{g_rho_q_q}) and (\ref{f_rho}) we get the
equation for the rho-quark coupling $g_{\rho qq}$
\begin{align*}
g_{\rho qq}^{-2} &  =-\frac{1}{8\pi^{2}}\left\{  \frac{1}{3}+\left(  m_{\rho
}^{2}+2M^{2}\right)  \left[  Y_{1}\left(  m_{\rho}^{2},M^{2}\right)  \right]
\right. \\
&  \ \ \ \ \ \ \ \ \ \ \ \ \ \ \ \left.  +\left[  Z_{0}\left(  m_{\rho}%
^{2},M^{2};M^{2}\right)  \right]  -\frac{\pi^{2}}{3M^{2}}\left[  \Pi_{\left(
L\right)  }^{AA}\left(  0\right)  \right]  \right\}  ,
\end{align*}
and the $f_{\rho}$%
\[
f_{\rho}=\frac{4G_{V}m_{\rho}^{2}}{g_{\rho qq}}.
\]
In order to find the axial-vector meson parameters we evaluated axial-axial
polarization function at $p^{2}=m_{a_{1}}^{2}$
\begin{align*}
\Pi_{\left(  T\right)  }^{AA}\left(  m_{a_{1}}^{2}\right)   &  =\frac{1}%
{6\pi^{2}}m_{a_{1}}^{2}+\frac{1}{2\pi^{2}}\left(  m_{a_{1}}^{2}-4M^{2}\right)
\left[  Z_{0}\left(  m_{a_{1}}^{2},M^{2};M^{2}\right)  \right] \\
&  +\left(  1-\frac{m_{a_{1}}^{2}}{6M^{2}}\right)  \left[  \Pi_{\left(
L\right)  }^{AA}\left(  0\right)  \right]  ,
\end{align*}
and obtain the $a_{1}^{0}$ mass condition:%
\[
m_{a_{1}}^{2}=-\frac{2\pi^{2}}{G_{V}}\left\{  \frac{1}{3}+\left(
1-\frac{4M^{2}}{m_{a_{1}}^{2}}\right)  \left[  Z_{0}\left(  m_{a_{1}}%
^{2},M^{2};M^{2}\right)  \right]  +2\pi^{2}\left(  \frac{1}{m_{a_{1}}^{2}%
}-\frac{1}{6M^{2}}\right)  \left[  \Pi_{\left(  L\right)  }^{AA}\left(
0\right)  \right]  \right\}  ^{-1},
\]
and also the $a_{1}$-quark coupling $g_{a_{1}qq}$%
\begin{align*}
g_{a_{1}qq}^{-2} &  =\frac{1}{8\pi^{2}}\left\{  \frac{1}{3}+\left(  m_{a_{1}%
}^{2}-6M^{2}\right)  \left[  Y_{1}\left(  m_{a_{1}}^{2},M^{2};M^{2}\right)
\right]  \right. \\
&  \ \ \ \ \ \ \ \ \ \ \ \ \ \ \ \left.  +\left[  Z_{0}\left(  m_{a_{1}}%
^{2},M^{2};M^{2}\right)  \right]  -\frac{\pi^{2}}{3M^{2}}\left[  \Pi_{\left(
L\right)  }^{AA}\left(  0\right)  \right]  \right\}  .
\end{align*}
We see that the remaining meson parameters are dependent of the vector
coupling. Adopting the $G_{V}\simeq16.4\;\mathrm{GeV}^{-2},$ as our last input
parameter, we can fix all the remaining quantities of the model. In this way
we obtain the masses of the vector and the axial-vector mesons, $m_{\rho
}\simeq770.0\;\mathrm{MeV}$ and $m_{a_{1}}\simeq1145.0\;\mathrm{MeV,}$ and
their corresponding effective coupling constants, $g_{\rho qq}\simeq4.8$ and
$g_{a_{1}qq}\simeq2.9,$ as well as $\sqrt{C_{crit}}\simeq0.96\;\mathrm{GeV}$
and $f_{\rho}\simeq8.0$.\ Again the values are in good accordance with the
expectations. In our description the meson $\rho$ is a bound state while the
meson $a_{1}$, as well known, are not bound state which implies that its mass
has imaginary part. This means that the description of the axial-vector mesons
within the NJL model are less realistic than that for the pseudoscalar and
vector mesons. In the above results we have discarded the imaginary parts of
results for $m_{a_{1}}$ and $g_{a_{1}qq}$. In the pion sector, the pion
coupling constant $\widetilde{g}_{\pi qq}$, which is dependent on $G_{V}$,
acquires the value $\widetilde{g}_{\pi qq}\simeq2.9$.

An interesting point is the one relative to the value found for $C_{crit}$. It
is very similar to that obtained for the cutoff parameter in traditional
treatments, which is usually located in the range $600-1000\;\mathrm{MeV}$. A
simple analysis reveals that this similarity is, in fact, expected. If we
remember that the constant $C$ has introduced in the general expression for
the quadratic divergence as an arbitrary constant, which means independence of
the scale parameter in the basic quadratic divergence $I_{quad}\left(
\lambda^{2}\right)  $, it is simple to see that it can be associated to the
dominant term ($\Lambda^{2}$) in the regularized version of the $I_{quad}%
\left(  \lambda^{2}\right)  $. Due to this reason the value for the $C_{crit}$
is expected to be closely related to the regularization parameter $\Lambda
^{2}$ of the proper-time and to the cutoff $\Lambda^{2}$ of the sharp cutoff regularization.

A similar analysis can be made about the values found for the quark mass $M$.
Although we have concluded that the unique acceptable physical prediction for
the mass is that dictated by the critical condition of the Eq.
(\ref{algebraic}), we could note that the values for the solutions $M_{1}$ and
$M_{3}$, and their corresponding partner $C$, are in agreement with the values
usually found in the literature. In fact, by choosing the value of $C$ we can
obtain a correspondence $(C,M)$ with those $(\Lambda,M)$ frequently presented
in the literature, in investigations made within the context of
regularizations. This aspect can be clearly observed in the figure (1). Note
that above the critical point, for each value of $C$, there are two
independent solutions, namely $M_{1}$ and $M_{3}$. The $M_{1}$ solutions grows
up with the increase of $C$ while $M_{3}$ asymptotically goes to zero.%
\begin{figure}
[h]
\begin{center}
\includegraphics[
natheight=5.737000in,
natwidth=5.448700in,
height=4.0739in,
width=3.8708in
]%
{MSNYEU01.gif}%
\end{center}
\end{figure}

In order to give a further clarification relative to these aspects we present,
in table I, some typical values for $M$ found in the literature obtained in
four representative regularizations schemes (see for example Ref.
\cite{Klevansky}), the values of $\Lambda$ (cutoff parameter) and the
associated value for the parameter $C$. We see that both $\Lambda$ and $C$ are
always of the same order but for the four-momentum cutoff and proper time
schemes the values are strictly the same.

\begin{center}%
\begin{tabular}
[c]{|c|c|c|c|}\hline
scheme & $M\;(MeV)$ & $\Lambda\;(MeV)$ & $\sqrt{C}\;(MeV)$\\\hline
three-momentum cutoff & $313$ & $653$ & $932$\\\hline
four-momentum cutoff & $238$ & $1015$ & $1017$\\\hline
proper time & $200$ & $1086$ & $1086$\\\hline
Pauli-Villars & $241$ & $859$ & $1012$\\\hline
\end{tabular}

\end{center}

In the next \ step we include in the discussion a nonvanishing current quark
mass ($m_{0}\neq0$) which gives to the pion a nonvanishing mass. Our first
task in this case is to found the new parametrization for the $I_{\log}\left(
M^{2}\right)  $ since the parametrization for $I_{quad}\left(  M^{2}\right)  $
does not change. In this direction we first note that the determinant $D_{\pi
}\left(  p^{2}\right)  $ becomes
\[
D_{\pi}\left(  p^{2}\right)  =\frac{m_{0}}{M}\left[  1+G_{V}\Pi_{\left(
L\right)  }^{AA}\left(  p^{2}\right)  \right]  -\frac{G_{S}}{4M^{2}}%
p^{2}\left[  \Pi_{\left(  L\right)  }^{AA}\left(  p^{2}\right)  \right]  ,
\]
where we have used the gap equation. The condition (\ref{m_pi_def}) plus
equation above determines the expression for the pion mass:%
\[
m_{\pi}^{2}=\frac{4m_{0}M}{G_{S}}\frac{\left[  1+G_{V}\Pi_{\left(  L\right)
}^{AA}\left(  m_{\pi}^{2}\right)  \right]  }{\left[  \Pi_{\left(  L\right)
}^{AA}\left(  m_{\pi}^{2}\right)  \right]  }.
\]
On the other hand the pion decay constant may be written as%
\[
f_{\pi}g_{\pi qq}^{-1}=\frac{1}{16M^{2}}\frac{\left[  \Pi_{\left(  L\right)
}^{AA}\left(  m_{\pi}^{2}\right)  \right]  }{\left[  1+G_{V}\Pi_{\left(
L\right)  }^{AA}\left(  m_{\pi}^{2}\right)  \right]  },
\]
where the coupling $g_{\pi qq}$ is given by%
\[
g_{\pi qq}^{-2}=\frac{\left[  \Pi_{\left(  L\right)  }^{AA}\left(  m_{\pi}%
^{2}\right)  \right]  }{4M^{2}\left[  1+G_{V}\Pi_{\left(  L\right)  }%
^{AA}\left(  m_{\pi}^{2}\right)  \right]  }-\frac{N_{c}}{\pi^{2}}\frac{m_{0}%
M}{G_{S}}\frac{\left[  Y_{1}\left(  m_{\pi}^{2},M^{2}\right)  \right]
}{\left[  \Pi_{\left(  L\right)  }^{AA}\left(  m_{\pi}^{2}\right)  \right]
\left[  1+G_{V}\Pi_{\left(  L\right)  }^{AA}\left(  m_{\pi}^{2}\right)
\right]  }.
\]
These equations together lead us to the expression%
\[
\frac{\left[  \Pi_{\left(  L\right)  }^{AA}\left(  m_{\pi}^{2}\right)
\right]  ^{3}}{\left[  1+G_{V}\Pi_{\left(  L\right)  }^{AA}\left(  m_{\pi}%
^{2}\right)  \right]  }-4f_{\pi}^{2}\left[  \Pi_{\left(  L\right)  }%
^{AA}\left(  m_{\pi}^{2}\right)  \right]  ^{2}+\frac{16N_{c}}{\pi^{2}}%
\frac{m_{0}M^{3}f_{\pi}^{2}}{G_{S}}\left[  Y_{1}\left(  m_{\pi}^{2}%
,M^{2}\right)  \right]  =0,
\]
whose solution gives the searched parametrization for $I_{\log}\left(
M^{2}\right)  $. Among the many solutions of this cubic algebraic equation
there is just one which satisfy the physical conditions expected for $I_{\log
}\left(  M^{2}\right)  $. That solution may be written as%
\[
\Pi_{\left(  L\right)  }^{AA}\left(  m_{\pi}^{2}\right)  =\frac{4}{3}%
\frac{f_{\pi}^{2}}{1-4G_{V}f_{\pi}^{2}}\left\{  1+\sqrt[3]{R-\sqrt{R^{2}%
-Q^{3}}}+\sqrt[3]{R+\sqrt{R^{2}-Q^{3}}}\right\}  ,
\]
where%
\begin{align*}
R  & =1-\frac{9N_{c}}{8\pi^{2}}\frac{m_{0}M^{3}}{G_{S}f_{\pi}^{4}}\left(
1-4G_{V}f_{\pi}^{2}\right)  \left(  3-8G_{V}f_{\pi}^{2}\right)  \left[
Y_{1}\left(  m_{\pi}^{2},M^{2}\right)  \right]  ,\\
Q  & =1-\frac{3N_{c}}{\pi^{2}}\frac{m_{0}M^{3}G_{V}}{G_{S}f_{\pi}^{2}}\left(
1-4G_{V}f_{\pi}^{2}\right)  \left[  Y_{1}\left(  m_{\pi}^{2},M^{2}\right)
\right]  .
\end{align*}
This gives the searched parametrization for $I_{\log}\left(  M^{2}\right)  $
\begin{align}
i\left[  I_{\log}\left(  M^{2}\right)  \right]    & =-\frac{1}{16\pi^{2}%
}\left[  Z_{0}\left(  m_{\pi}^{2},M^{2};M^{2}\right)  \right]  \nonumber\\
& -\frac{1}{12N_{c}M^{2}}\frac{f_{\pi}^{2}}{1-4G_{V}f_{\pi}^{2}}\left\{
1+\sqrt[3]{R-\sqrt{R^{2}-Q^{3}}}+\sqrt[3]{R+\sqrt{R^{2}-Q^{3}}}\right\}
.\label{I_log_2}%
\end{align}
Replacing the parametrizations (\ref{I_log_2}) and (\ref{I_quad}) in
(\ref{rel_3}) we get%
\[
M^{3}\left(  1+\left[  Z_{0}\left(  m_{\pi}^{2},M^{2};M^{2}\right)  \right]
\right)  +\left(  \frac{\pi^{2}}{3}\left[  \Pi_{\left(  L\right)  }%
^{AA}\left(  m_{\pi}^{2}\right)  \right]  -C\right)  M-\frac{4\pi^{2}}%
{3}\left\langle \overline{\psi}\psi\right\rangle =0.
\]
This is a nontrivial nonlinear equation for $M$ which may be simplified by
using the following\ (reasonable) approximations%
\begin{align*}
Z_{0}\left(  m_{\pi}^{2},M^{2};M^{2}\right)    & \simeq-\frac{m_{\pi}^{2}%
}{6M^{2}},\\
Y_{1}\left(  m_{\pi}^{2},M^{2}\right)    & \simeq-\frac{1}{6M^{2}},
\end{align*}
which, among others things, allow us to see clearly the searched solutions for
$M$. Then we get%
\begin{equation}
M^{3}+\left(  \frac{\pi^{2}}{3}\left[  \Pi_{\left(  L\right)  }^{AA}\right]
+\frac{4}{3}m_{0}\left\langle \overline{\psi}\psi\right\rangle \frac{\left[
1+G_{V}\Pi_{\left(  L\right)  }^{AA}\right]  }{\left[  \Pi_{\left(  L\right)
}^{AA}\right]  }-C\right)  M-\frac{4\pi^{2}}{3}\left\langle \overline{\psi
}\psi\right\rangle =0,\label{cubic_eq}%
\end{equation}
where we have also used that%
\begin{align*}
m_{\pi}^{2}  & \simeq-8m_{0}\left\langle \overline{\psi}\psi\right\rangle
\frac{\left[  1+G_{V}\Pi_{\left(  L\right)  }^{AA}\right]  }{\left[
\Pi_{\left(  L\right)  }^{AA}\right]  },\\
R  & \simeq1-\frac{3N_{c}}{8\pi^{2}}\frac{m_{0}\left\langle \overline{\psi
}\psi\right\rangle }{f_{\pi}^{4}}\left(  1-4G_{V}f_{\pi}^{2}\right)  \left(
3-8G_{V}f_{\pi}^{2}\right)  ,\\
Q  & \simeq1-\frac{N_{c}}{\pi^{2}}\frac{m_{0}\left\langle \overline{\psi}%
\psi\right\rangle }{f_{\pi}^{2}}G_{V}\left(  1-4G_{V}f_{\pi}^{2}\right)  .
\end{align*}
The equation (\ref{cubic_eq}) has the form $M^{3}+\alpha M+\beta=0$ where the
coefficients $\alpha$ and $\beta$ are given by%
\begin{align*}
\alpha & =\frac{\pi^{2}}{3}\left[  \Pi_{\left(  L\right)  }^{AA}\right]
+\frac{4}{3}m_{0}\left\langle \overline{\psi}\psi\right\rangle \frac{\left[
1+G_{V}\Pi_{\left(  L\right)  }^{AA}\right]  }{\left[  \Pi_{\left(  L\right)
}^{AA}\right]  }-C,\\
\beta & =-\frac{4\pi^{2}}{3}\left\langle \overline{\psi}\psi\right\rangle .
\end{align*}
Then, in order to find its solutions we follows strictly the same steps which
we have used for the Eq.(\ref{algebraic}). The critical solution does not
change (see Eq.(\ref{M_c})) while the critical value of the constant $C$
become%
\[
C_{crit}=\frac{\pi^{2}}{3}\left[  \Pi_{\left(  L\right)  }^{AA}\right]
+\frac{4}{3}m_{0}\left\langle \overline{\psi}\psi\right\rangle \frac{\left[
1+G_{V}\Pi_{\left(  L\right)  }^{AA}\right]  }{\left[  \Pi_{\left(  L\right)
}^{AA}\right]  }+\sqrt[3]{12\pi^{4}\left\langle \overline{\psi}\psi
\right\rangle ^{2}}.
\]
The noncritical solutions $M_{1}$ and $M_{2}$ are given by Eqs. (\ref{M1}) and
(\ref{M2}) with
\begin{align*}
A  & =2\sqrt{\frac{C}{3}-\frac{\pi^{2}}{9}\left[  \Pi_{\left(  L\right)
}^{AA}\right]  -\frac{4}{9}m_{0}\left\langle \overline{\psi}\psi\right\rangle
\frac{\left[  1+G_{V}\Pi_{\left(  L\right)  }^{AA}\right]  }{\left[
\Pi_{\left(  L\right)  }^{AA}\right]  }},\\
\cos\theta & =\frac{6\pi^{2}\left\langle \overline{\psi}\psi\right\rangle
}{\sqrt{3\left(  C-\frac{\pi^{2}}{3}\left[  \Pi_{\left(  L\right)  }%
^{AA}\right]  -\frac{4}{3}m_{0}\left\langle \overline{\psi}\psi\right\rangle
\left(  G_{V}+\left[  \Pi_{\left(  L\right)  }^{AA}\right]  ^{-1}\right)
\right)  ^{3}}}.
\end{align*}
These solutions when plotted as function of $C$ give a graphic similar to that
shown in the Fig. (1).

In the vector sector, the $\rho$ mass and its coupling constant are given,
respectively, by
\begin{align*}
m_{\rho}^{-2}  & =-\frac{G_{V}}{2\pi^{2}}\left\{  \frac{1}{3}+\left(
1+\frac{2M^{2}}{m_{\rho}^{2}}\right)  \left[  Z_{0}\left(  m_{\rho}^{2}%
,M^{2};M^{2}\right)  \right]  \right. \\
& \ \ \ \ \ \ \ \ \ \ \left.  -\left[  Z_{0}\left(  m_{\pi}^{2},M^{2}%
;M^{2}\right)  \right]  -\frac{\pi^{2}}{3M^{2}}\left[  \Pi_{\left(  L\right)
}^{AA}\left(  m_{\pi}^{2}\right)  \right]  \right\}  ,
\end{align*}
and%
\begin{align*}
g_{\rho qq}^{-2} &  =-\frac{1}{24\pi^{2}}-\frac{1}{8\pi^{2}}\left(  m_{\rho
}^{2}+2M^{2}\right)  \left[  Y_{1}\left(  m_{\rho}^{2},M^{2}\right)  \right]
\\
&  -\frac{1}{8\pi^{2}}\left[  Z_{0}\left(  m_{\rho}^{2},M^{2};M^{2}\right)
-Z_{0}\left(  m_{\pi}^{2},M^{2};M^{2}\right)  \right] \\
&  +\frac{1}{24M^{2}}\left[  \Pi_{\left(  L\right)  }^{AA}\left(  m_{\pi}%
^{2}\right)  \right]  ,
\end{align*}
while for the $a_{1}$ meson the results are%
\begin{align*}
m_{a_{1}}^{-2} &  =-\frac{N_{c}G_{V}}{6\pi^{2}}\left\{  \frac{1}{3}+\left(
1-\frac{4M^{2}}{m_{a_{1}}^{2}}\right)  \left[  Z_{0}\left(  m_{a_{1}}%
^{2},M^{2};M^{2}\right)  \right]  \right. \\
\ \ \ \  &  \ \ \ \ \ \ \ \ \ \ \ \ \ \ \ \ \ \left.  -\left(  1-\frac{6M^{2}%
}{m_{a_{1}}^{2}}\right)  \left[  Z_{0}\left(  m_{\pi}^{2},M^{2};M^{2}\right)
\right]  +\frac{6\pi^{2}}{N_{c}}\left(  \frac{1}{m_{a_{1}}^{2}}-\frac
{1}{6M^{2}}\right)  \left[  \Pi_{\left(  L\right)  }^{AA}\left(  m_{\pi}%
^{2}\right)  \right]  \right\}  ,
\end{align*}
and%
\begin{align*}
g_{a_{1}qq}^{-2} &  =\frac{N_{c}}{72\pi^{2}}+\frac{N_{c}}{24\pi^{2}}\left(
m_{a_{1}}^{2}-4M^{2}\right)  \left[  Y_{1}\left(  m_{a_{1}}^{2},M^{2}\right)
\right] \\
&  +\frac{N_{c}}{24\pi^{2}}\left[  Z_{0}\left(  m_{a_{1}}^{2},M^{2}%
;M^{2}\right)  -Z_{0}\left(  m_{\pi}^{2},M^{2};M^{2}\right)  \right] \\
&  -\frac{1}{24M^{2}}\left[  \Pi_{\left(  L\right)  }^{AA}\left(  m_{\pi}%
^{2}\right)  \right]  .
\end{align*}
Now we are ready to obtain the numerical values for the parameter which belong
to the meson phenomenology, like we did for the chirally symmetric case, and
to compare them with the ones obtained both in literature of this issue and in
the experiments. Again we assume $\left\langle \overline{\psi}\psi
\right\rangle =\left(  -250.0\;\mathrm{MeV}\right)  ^{3}$, $f_{\pi
}=93.0\;\mathrm{MeV}$ and $G_{V}\simeq16.4\;\mathrm{GeV}^{-2}$ as our inputs
parameters. On the other hand, in the chirally nonsymmetric case we have added
a new parameter in the Lagrangian which break explicit the chiral symmetry,
the current quark mass $m_{0}$. We consider the $m_{0}$ as our last input
parameter with the value $m_{0}=5.1\ \mathrm{MeV}$. We now obtain a
nonvanishing pion mass with the value $m_{\pi}=136.7\;\mathrm{MeV}$ which are
in good agreement with the experimental one. Concerning with the remaining
parameters, we have verified that the results obtained for the case of
$m_{0}=0$ does not change appreciably. Of course this is due to the fact that
our constituent quark mass $M$ does not change when $m_{0}\neq0$. When we
compare our results with the ones predicted by the experiments we see that
they are globally very good. Its gratifying for us to see that a consistent
treatment of the divergencies when applied to a nonrenormalizable model, in
the same way what is done for renormalizable theories, furnish good
predictions for the phenomenological observables. It is important to emphasize
one more time that the number of input parameters are precisely the ones
present in the model Lagrangean and that results are completely independent of
the intrinsic arbitrariness or choices made in intermediary steps of this type
of calculations. In addition to these very attractive features the results
obtained are in excellent agreement with the experimental values.

As a last comment we note that when $G_{V}=0$, the results obtained in this
work give the same ones produced by our previous work \cite{Orimar-PRD-NJL},
as should be expected. In particular, in the pion sector analytical results
may be written as%
\begin{equation}
m_{\pi}^{2}=\sqrt[3]{\delta_{1}-\delta_{2}}-\sqrt[3]{\delta_{1}+\delta_{2}},
\end{equation}%
\begin{equation}
g_{\pi qq}^{-2}=\frac{N_{c}N_{f}m_{\pi}^{2}}{48\pi^{2}\sqrt[3]{\left(
\frac{2\pi^{2}}{N_{c}}\left\langle \overline{\psi}\psi\right\rangle \right)
^{2}}}+\frac{f_{\pi}^{2}}{2\sqrt[3]{\left(  \frac{2\pi^{2}}{N_{c}}\left\langle
\overline{\psi}\psi\right\rangle \right)  ^{2}}}\left(  1+\sqrt{1+\frac
{N_{c}N_{f}m_{\pi}^{2}}{12\pi^{2}f_{\pi}^{2}}}\right)  ,
\end{equation}
where
\begin{align*}
\delta_{1}  & =\sqrt{-\frac{128\pi^{6}N_{f}m_{0}^{3}\left\langle
\overline{\psi}\psi\right\rangle ^{4}}{N_{c}^{3}\left(  m_{0}-\sqrt[3]%
{-\frac{2\pi^{2}}{N_{c}}\left\langle \overline{\psi}\psi\right\rangle
}\right)  ^{3}}\left(  32\gamma+\frac{9N_{f}m_{0}\sqrt[3]{-\frac{2\pi^{2}%
}{N_{c}}\left\langle \overline{\psi}\psi\right\rangle }\left\langle
\overline{\psi}\psi\right\rangle }{\left(  m_{0}-\sqrt[3]{-\frac{2\pi^{2}%
}{N_{c}}\left\langle \overline{\psi}\psi\right\rangle }\right)  f_{\pi}^{4}%
}\right)  },\\
\delta_{2}  & =-\frac{\frac{6\pi^{2}}{N_{c}N_{f}}m_{0}^{2}\sqrt[3]{\left(
\frac{2\pi^{2}}{N_{c}}\left\langle \overline{\psi}\psi\right\rangle \right)
^{2}}\left(  2N_{f}\left\langle \overline{\psi}\psi\right\rangle \right)
^{2}}{f_{\pi}^{2}\left(  m_{0}-\sqrt[3]{-\frac{2\pi^{2}}{N_{c}}\left\langle
\overline{\psi}\psi\right\rangle }\right)  ^{2}}.
\end{align*}
which complete our calculations.

\section{Summary and Conclusions}

In the preceding sections, we considered the question of predictive power of
the NJL model. Traditionally the model has been used to describe low-energy
hadronic observables, in spite of its nonrenormalizable character and, for
this reason, the corresponding predictions have been constructed in a
previously assumed level of approximation and compromising with a particular
regularization prescription. Within this context, it is well known that the
results, for the evaluated model amplitudes, invariably emerges as dependent
on many types of choices made in intermediary steps of the calculations. The
first and immediate of such choices is the regularization prescription. Since
in nonrenormalizable models the regularization cannot be removed, the results
are assumed regularization dependent. In addition, due to the fact that the DR
is not adequate for the treatment of the model amplitudes, the results may
emerge as dependent on choices for the routing of internal lines momenta in
loops or dependent on arbitrary scales used in the separation of terms having
different degrees of divergences. The ambiguous terms, on the other hand, are
invariable associated with the violation of symmetries implemented in the
model construction. In this scenario, it becomes difficult to talk about the
model predictions. Among other aspects, the model is overparametrized just
because at least one regularization parameter needs to be specified. This
means that one observable, which is in the scope of the model predictions,
must be used as an additional input in order to parametrize the model. Given
this situation, a relevant question can be put: is it possible to make genuine
predictions within the NJL model or the facts described above are definitive?

Having this question in mind, we proposed a very general investigation
involving all one and two-point function of the SU(2) version of the NJL model
\cite{Orimar-PRD-NJL}. The conclusion stated reveals the possibility of making
the predictions free of ambiguities or symmetry violating terms as well as
free from the dependence on the specific regularization choice. This
conclusion was made possible just because the calculations were performed
following a novel strategy of handling the divergences in the perturbative
calculations of QFT. Within this strategy, divergent integrals are never
really evaluated. Only general properties for standard divergent objects are
adopted which are dictated by the consistent and systematic elimination of
ambiguous and symmetry violating terms. The surprising fact that emerged in
the above cited investigation, which is particular to the NJL model, refers to
a critical condition obtained in the equation stating the dynamically
generated quark mass. After imposing all the constraints coming from the
consistency of the perturbative calculations, the existence of freedom
associated with the specific form of the regularization was observed. However,
when the constituent quark mass is searched for, as a function of the model
input parameters as well as a function of the arbitrary parameter representing
the arbitrariness remaining, it was observed that a unique reasonable physical
solution exists. Only one real value for the quark mass emerges through a
critical condition, which fixes the arbitrariness remaining. Since, after
this, the model predictions are completely independent on choices, this
formulation was denominated predictive.

Following this line of reasoning, the present work can be considered as an
additional step relative to Ref. \cite{Orimar-PRD-NJL}. Here we put the
formulation within a more general context and, through analytical solutions,
we show in a clear way the origin of the critical condition by obtaining the
expressions for the critical value of the arbitrary parameter involved as well
as the value of the corresponding constituent quark mass in analytical forms.
In addition, we show how to explain the values for the quark mass found in the
literature by using traditional regularization schemes. The model considered
here for the phenomenology predictions was also extended to consider also the
vector mesons. The relevant steps of the strategy adopted in the reported
investigation to handle the divergences, which allowed our conclusions, can be
summarized as follows:

i) Identify the amplitudes, pertinent to the model, having the highest
superficial degree of divergence $D$ ($D=3$ for the present case).

ii) Through the expression (\ref{fermion_prop}) specify the convenient
representation for the involved propagators, taken $N=D$ and making the
summation indicated. For the present case this means to adopt the
representation (\ref{fermion_prop2}).

iii) Through the Feynman rules, construct the amplitudes, performing all the
operation like Dirac or other involved traces operation, algebraic
manipulation, convenient reorganizations etc..., except the integration in the
loop momentum.

iv) Introduce the integration over the loop momentum, performing the
integration in all finite Feynman integrals and letting all the divergent ones
unchanged. The obtained result can be put in terms of standard finite
functions and standard divergent objects. In the present case the functions
(\ref{Z_k}) and the objects (\ref{I_log_def}), (\ref{I_quad_def}), and
(\ref{Delta}).

v) Remove the divergent objects which are differences among divergent
integrals of the same divergences degree guided by the maintenance of
fundamental symmetries like the space-time homogeneity in perturbative
calculations, which we denominate consistency relations, defining thus the
consistent regularized amplitudes. In the present case this means just to
require $\Delta_{\alpha\beta}\left(  \lambda^{2}\right)  =$ $\nabla
_{\alpha\beta}\left(  \lambda^{2}\right)  =\square_{\alpha\beta\mu\nu}\left(
\lambda^{2}\right)  =0$.

vi) Impose the scale independence over the full amplitudes (finite and
divergent parts) stating then the scale properties for the irreducible
divergent objects. In the present case this means to state the properties
(\ref{scale_1}) and (\ref{scale_2}) which are relations among the irreducible
divergent objects at different scales. Such relations imply definite
properties for the irreducible divergent objects, which are for the present
case shown in Eqs. (\ref{C1}) and (\ref{rel_3}). These properties will state
relations among the divergent objects which allow us to reduce the freedom
remaining to a unique arbitrary constant.

vii) The remaining divergent objects must be removed through the
reparametrization of the model at the considered level of approximation. This
means to relate them to the physical inputs of the model (similar to a
renormalization in renormalizable theories).

viii) To find the best physical values for the remaining arbitrariness. This
last step will obviously depend on the specific problem. In the NJL model, as
we have shown, this search for the best physical value has ended in the
existence of a critical condition. Applications of this strategy to the SU(3)
case have been considered revealing the same conclusion stated here: within
the context of the adopted strategy to handle the divergences the NJL model
becomes predictive. Work along this line is presently under way.

\end{document}